\documentclass[12pt]{article}

\input{epsf}

\begin{document}

\begin{center}

{\Large\bf Nuclear collective motion with a coherent coupling
interaction between quadrupole and octupole modes}\\
\bigskip\bigskip

N. Minkov$^\dagger$\footnote{E-mail: nminkov@inrne.bas.bg},
P. Yotov$^\dagger$, S. Drenska$^\dagger$, W. Scheid$^+$,
D. Bonatsos$^{\#}$, D. Lenis$^{\#}$, and D. Petrellis$^{\#}$\\
\medskip

{\em $^\dagger$ Institute of Nuclear Research and Nuclear Energy,\\
72 Tzarigrad Road, 1784 Sofia, Bulgaria} \\
\medskip

{\em $^+$ Institut f\"{u}r Theoretische Physik der
Justus-Liebig-Universit\"at,\\ Heinrich-Buff-Ring 16, D--35392
Giessen, Germany}
\medskip

{\em $^{\#}$ Institute of Nuclear Physics, N.C.S.R. ``Demokritos'',\\
GR--15310 Aghia Paraskevi, Attiki, Greece}

\bigskip\bigskip

{\bf Abstract}

\end{center}

A collective Hamiltonian for the rotation--vibration motion of
nuclei is considered, in which the axial quadrupole and octupole
degrees of freedom are coupled through the centrifugal interaction.
The potential of the system depends on the two deformation variables
$\beta_2$ and $\beta_3$. The system is considered to oscillate
between positive and negative $\beta_3$-values, by rounding an
infinite potential core in the $(\beta_2,\beta_3)$-plane with
$\beta_2>0$. By assuming a coherent contribution of the quadrupole
and octupole oscillation modes in the collective motion, the energy
spectrum is derived in an explicit analytic form, providing
specific parity shift effects. On this basis several possible ways
in the evolution of quadrupole--octupole collectivity are outlined.
A particular application of the model to the energy levels and
electric transition probabilities in alternating parity spectra of
the nuclei $^{150}$Nd, $^{152}$Sm, $^{154}$Gd and $^{156}$Dy is
presented.

\medskip

PACS: 21.60.Ev; 21.10.Re
\medskip

%Section: Nuclear Structure

\newpage

\section{Introduction}

Shape deformations and surface oscillations in atomic nuclei
determine from a geometric point of view the main features of
nuclear collective dynamics \cite{BM75}. The leading quadrupole
mode manifests itself in all regions of collectivity providing
vibrational, rotational, and transitional structures of the spectra. In
addition, in some regions the manifestation of octupole degrees
of freedom is superposed, leading to more complicated shape
properties and parity effects in the spectrum of the system
\cite{AB,BN96}. A variety of microscopic, geometric and algebraic
model approaches have been applied in nuclear regions where the
quadrupole and octupole degrees of freedom coexist \cite{BN96}.

In general, the problem of quadrupole--octupole collectivity is not
easy to solve neither microscopically, mainly due to the breaking
of reflection symmetry, nor geometrically, due to the difficulty in
determining the total inertia tensor of the system. It is,
however, simplified considerably if the axial symmetry is still
preserved and if the octupole deformations are fixed appropriately
with respect to the principal axes of the quadrupole shape.
Further simplification is achieved if both degrees of freedom are
separated adiabatically. It allows one to examine the manifestation
of the octupole mode for fixed values of quadrupole parameters. In
such a case the collective motion can be associated to the
oscillations of the reflection asymmetric shape with respect to an
octupole variable $\beta_3$ in a double-well potential
\cite{Krappe69,Leander}. Then the parity shift effect observed in
nuclear alternating parity bands can be explained as the result of
the tunnelling through the potential barrier \cite{Jolos93,Jolos94}.
This concept has been generalized for the case of simultaneously
contributing quadrupole and octupole modes \cite{DD93}, as well as
for the case of higher multipole degrees of freedom \cite{DD95}.
In both cases the double-well potential was defined in terms of
a variable carrying the relative contribution of the different
degrees of freedom and not the absolute values of the respective
deformation variables. In such a way the explicit form of the
original potential in terms of the quadrupole $\beta_2$ and octupole
$\beta_3$ deformation variables was not given. As a consequence, some
basic characteristics of the quadrupole and octupole modes and their
interaction remain outside of consideration. Such is the behavior of
the system in dependence on the quadrupole and octupole stiffness, as
well as the limiting case of a frozen quadrupole variable. Another
interesting question is, if and to what extent one may consider the
presence of a tunnelling effect in the space of the octupole variable
$\beta_3$ after the quadrupole coordinate $\beta_2$ is let to vary.
Some limiting cases in the shape evolution and the angular momentum
properties of the system are also of interest in respect with the
above.

The purpose of the present work is to clarify the above questions
by applying a simple explicit form of the collective energy
potential as a function of the quadrupole and octupole axial
deformation variables $\beta_2$ and $\beta_3$. We examine the
evolution of the potential shape in dependence on both degrees
of freedom, as well as on the collective angular momentum. The
geometric analysis suggests that the oscillations of the system in
the two-dimensional case of simultaneous manifestation of the quadrupole
and octupole modes are performed in a different way, compared to the
one-dimensional case of a reflection asymmetric shape with a frozen
quadrupole variable. We study the physical consequences of the
two-dimensional oscillations and demonstrate their role in the
rotation-vibration motion of the system.

In particular, the explicit geometric analysis of the
quadrupole--octupole potential suggests a possibility for a coherent
interplay between both collective modes. This allows the derivation
of explicit analytic expressions for the energy levels and
electromagnetic transition probabilities applicable to nuclei in which
an ``equal'' (coherent) manifestation of quadrupole and octupole degrees
of freedom is considered. As a result, one is able to study in detail the
respective effects in the structure of the spectrum. Below it
will be shown that such a consideration can be applied reasonably
to some nuclei in the rare earth region, such as the $N=90$ isotones
$^{150}$Nd, $^{152}$Sm, $^{154}$Gd and $^{156}$Dy. These nuclei are also
a subject of interest \cite{CZ,ClarkX5} from the point of view of the X(5)
critical point symmetry \cite{IacX5} between quadrupole vibrations [U(5)] and
axial quadrupole deformation [SU(3)]. In the present work we shall,
however, mainly consider the common quadrupole-octupole collective
properties, which, in principle, can take place in various nuclear regions.

In Sec. 2 the Hamiltonian of the coupled quadrupole and octupole modes
is presented, together with the geometric analysis of the
quadrupole--octupole potential. In Sec. 3 the Schr\"{o}dinger equation
is considered in the case of a coherent interplay between the two degrees
of freedom. The analytic solutions for several particular forms of the
potential and the respective schematic spectra are given in Sec. 4. In
addition, results of the model description of alternating parity spectra in
$^{150}$Nd, $^{152}$Sm, $^{154}$Gd and $^{156}$Dy are presented. The
electric transition probabilities are considered in Sec. 5, while in Sec. 6
a brief discusssion of the influence of the $\gamma$ degree of freedom
on the present results is given. Finally, a summary and concluding remarks
are given in Sec.~7.

\section{Hamiltonian for the coupled quadrupole and octupole modes}

We assume that the system is allowed to oscillate with respect to
the quadrupole $\beta_2$  and octupole $\beta_3$ axial deformation
variables. In addition, both degrees of freedom are coupled through
a centrifugal (rotation-vibration) interaction depending on the
collective angular momentum $I$. The energy potential represents a
two-dimensional surface determined by the variables $\beta_2$ and
$\beta_3$.

The quadrupole--octupole Hamiltonian describing the collective
motion under the above assumptions has the form
\begin{eqnarray}
H_{qo}&=&-\frac{\hbar^2}{2B_2}\frac{\partial^2}{\partial\beta_2^2}
-\frac{\hbar^2}{2B_3}\frac{\partial^2}{\partial\beta_3^2}+
U(\beta_2,\beta_3,I) \ , \label{Hqo}
\end{eqnarray}
where the potential is
\begin{equation}
U(\beta_2,\beta_3, I)=\frac{1}{2}C_2{\beta_2}^{2}+
\frac{1}{2}C_3{\beta_3}^{2} +
\frac{X(I)}{d_2\beta_2^2+d_3\beta_3^2}, \label{Ub2b3I}
\end{equation}
with $X(I)=I(I+1)/2$. Here $B_2$ and $B_3$ are the effective
quadrupole and octupole mass parameters, and $C_2$ and $C_3$ are
the stiffness parameters for the respective oscillation modes.

The last term in Eq.~(\ref{Ub2b3I}) provides a coupling between
quadrupole and octupole degrees of freedom. Its denominator can be
associated to the moment of inertia of an axially symmetric
quadrupole-octupole deformed shape, ${\mathcal{J}}^{(quad+oct)}=
3B_2\beta_{2}^2+6B_3\beta_{3}^2$ \cite{JPDav68}. Therefore, the
constants $d_2,\, d_3>0$ can be related to the mass parameters as
$d_2=3B_2$ and $d_3=6B_3$.  However, in the present study we do
not impose this relation and below a more general correlation
between $d_2,\, B_2$ and $d_3,\, B_3$ is considered. The quantities
$d_2$ and $d_3$ determine the contributions of the quadrupole and
octupole modes, respectively, to the moment of inertia. Also, we
remark that if the ground state of the system is considered ($I=0$),
the potential $U(\beta_2,\beta_3 ,I )$ should be taken by replacing
$X(I)\rightarrow d_0+X(I)$, with $d_0$ being a constant.

The Hamiltonian (\ref{Hqo}) represents a two-dimensional
generalization of the soft octupole oscillator Hamiltonian
introduced in \cite{MYDS06}, as well as of the one-dimensional
octupole Hamiltonian derived in \cite{Bizz05}. In the latter two
approaches the quadrupole mode is assumed frozen as mentioned in
Sec. 1. In this respect, Eq.~(\ref{Hqo}) corresponds to an
extension in which the quadrupole coordinate is let to vary. Also,
it corresponds to the quadrupole--octupole Hamiltonian in Refs.
\cite{DD93,DD95}. However, in the present work the potential
energy (\ref{Ub2b3I}) is taken in an explicit form depending on
$\beta_2$ and $\beta_3$ (including the harmonic oscillator part),
while in Refs. \cite{DD93,DD95} a double-oscillator potential is
defined in the space of polar coordinates. In this way, the
explicit form of Eq.~(\ref{Ub2b3I}) allows one to examine in
detail the potential surface and its dependence on the model
parameters and the collective angular momentum.

\begin{figure}[t]
\epsfxsize=10.cm   %or \epsfysize= HEIGHT cm
\centerline{\epsfbox{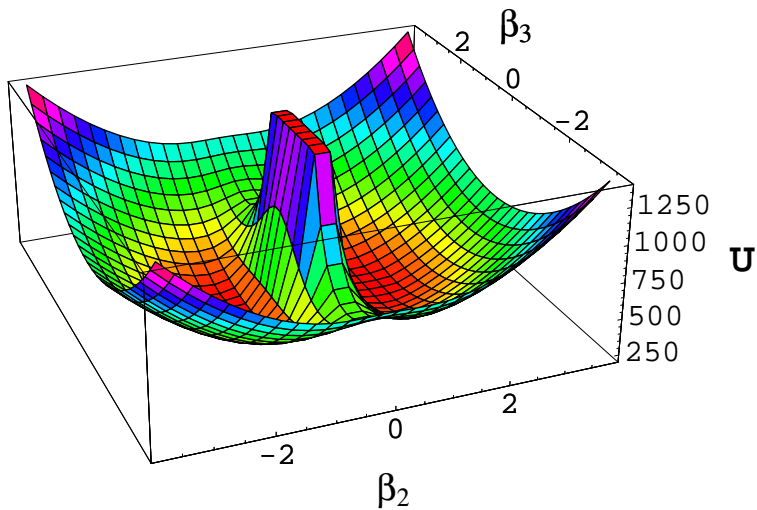}} \caption{(Color online) Schematic
3D plot of the potential $U(\beta_2,\beta_3,I)$,
Eq.~(\protect\ref{Ub2b3I}), in MeV for $I=5$ as a function of
$\beta_2$ and $\beta_3$. The parameter values are $C_2=C_3=100$
MeV, $d_2=0.1$ $\hbar^2$MeV$^{-1}$, and $d_3=0.01$
$\hbar^2$MeV$^{-1}$.} \label{fig:01}
\end{figure}
\begin{figure}[t]
\centerline{\epsfxsize=14.cm\epsfbox{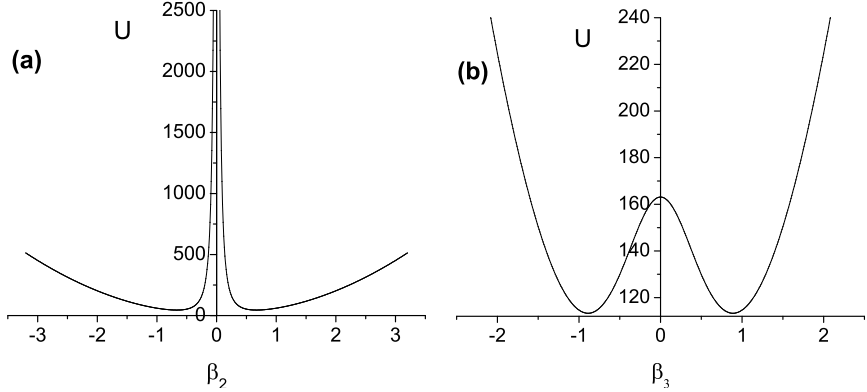}} \caption{Schematic
numerical behavior of the potential $U(\beta_2,\beta_3, I)$,
Eq.~(\protect\ref{Ub2b3I}), in MeV for $I=1$ as a function of: (a)
$\beta_2$ at fixed $\beta_3=0.1$; (b) $\beta_3$ at fixed
$\beta_2=0.25$. The parameter values are $C_2=C_3=100$ MeV,
$d_2=0.1$ $\hbar^2$MeV$^{-1}$, and $d_3=0.01$
$\hbar^2$MeV$^{-1}$.} \label{fig:02}
\end{figure}

Having in mind that the quadrupole deformation has the leading
role in the rotation mode, we assume that its contribution to the
moment of inertia is larger than the octupole contribution. This
assumption corresponds to the condition $d_2>d_3$, e.g. we can take
$d_2=0.1$ $\hbar^2$MeV$^{-1}$ and $d_3=0.01$ $\hbar^2$MeV$^{-1}$.
Then, for comparable values of the deformation variables $\beta_2$
and $\beta_3$, the input of the quadrupole mode in the denominator
of the centrifugal term will be larger than the octupole one. On
the other hand, as it will be seen in Sec. 3, this
circumstance does not restrict the possibility of equal (coherent)
contributions of both degrees of freedom in the mixed
quadrupole--octupole oscillation mode. Moreover, since the rotation
and vibration modes are coupled, the above condition might be not
strictly imposed. In this meaning the considered values of $d_2$
and $d_3$ provide only a schematic geometric analysis of the
potential (\ref{Ub2b3I}).

Let us now examine the minimum of the potential energy in dependence
on the model parameters. The set of extremum conditions for the
coordinates of the two-dimensional minimum
($\beta_{2min},\beta_{3min}$) is
\begin{eqnarray}
\left. \frac{\partial}{\partial
\beta_2}U(\beta_2,\beta_3,I) \right|_{(\beta_{2min},\beta_{3min})}=0\ \
\mbox{and}\ \
\left.\frac{\partial}{\partial \beta_3}U(\beta_2,\beta_3,I)
\right|_{(\beta_{2min},\beta_{3min})}=0;
\end{eqnarray}
\begin{eqnarray}
\left. \frac{\partial^2}{\partial
\beta_2^2}U(\beta_2,\beta_3,I) \right|_{(\beta_{2min},\beta_{3min})}>0\ \
\mbox{and}\ \
\left.\frac{\partial^2}{\partial \beta_3^2}U(\beta_2,\beta_3,I)
\right|_{(\beta_{2min},\beta_{3min})}>0.
\end{eqnarray}
It determines the following possible cases for the bottom of the
potential,
\smallskip

i) $\beta_{3min}=0$; $\beta_{2min}=\pm\left[ 2X(I)/(d_2C_2)
\right]^{1/4}$;
\smallskip

ii) $\beta_{2min}=0$; $\beta_{3min}=\pm\left[ 2X(I)/(d_3C_3)
\right]^{1/4}$;
\smallskip

iii) $\beta_{2min}\neq0$ and $\beta_{3min}\neq0$ with the
condition
\begin{eqnarray}
C_2=\frac{2X(I)d_2}{(d_2\beta_{2min}^2+d_3\beta_{3min}^2 )^2}
\ \
\mbox{and}\ \
C_3=\frac{2X(I)d_3}{(d_2\beta_{2min}^2+d_3\beta_{3min}^2)^2}\ .
\label{CL}
\end{eqnarray}

The shape of the potential corresponding to the case i) is
illustrated in Fig. 1. It is characterized by two energy minima
for $\beta_2>0$ and $\beta_2<0$ separated by a well determined
potential barrier. For given sign of $\beta_2$ (we consider
$\beta_2>0$) the bottom of the potential is not separated in the
$\beta_3$- direction, allowing oscillations of the system between
$\beta_3>0$ and $\beta_3<0$. This situation is illustrated in
Fig. 2. We see that for a fixed physically typical value of
$\beta_3$ (Fig.2a) the barrier in the quadrupole space of
$\beta_2$ is very large. Thus it restricts the values of the
quadrupole deformation within the half space $\beta_2>0$. For
a fixed typical $\beta_2$- value (Fig.2b) the barrier in the
octupole space of $\beta_3$ is relatively small. From Fig. 1 it
is seen that for some higher $\beta_2$- values this barrier
is reduced, and for $\beta_2\geq \beta_{2min}$ it disappears
($\beta_{3min}=0$).

In the case ii) the potential shape is the same as in Fig. 1, but
the coordinates $\beta_2$ and $\beta_3$ are exchanged. As far as
the system is not considered to oscillate between positive and
negative $\beta_2$ deformations, this case is not of interest in
the context of the present analysis.

\begin{figure}[t]
\epsfxsize=10.cm   %or \epsfysize= HEIGHT cm
\centerline{\epsfbox{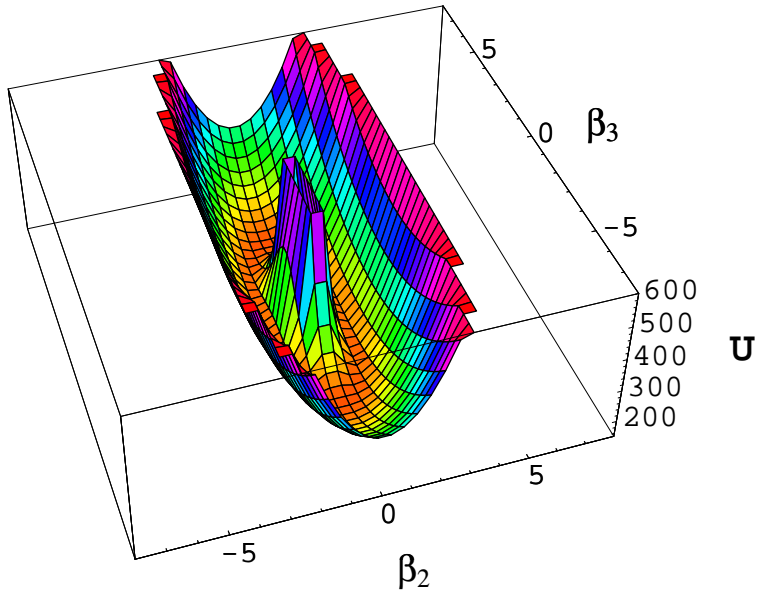}} \caption{(Color online) Schematic
3D plot of the potential $U(\beta_2,\beta_3,I)$,
Eq.~(\protect\ref{Ub2b3I}), in MeV for $I=5$ as a function of
$\beta_2$ and $\beta_3$. The parameter values are $C_2=100$ MeV,
$C_3=10$ MeV $d_2=0.1$ $\hbar^2$MeV$^{-1}$, and $d_3=0.01$
$\hbar^2$MeV$^{-1}$.} \label{fig:03}
\end{figure}
\begin{figure}[t]
\centerline{{\epsfxsize=8.cm\epsfbox{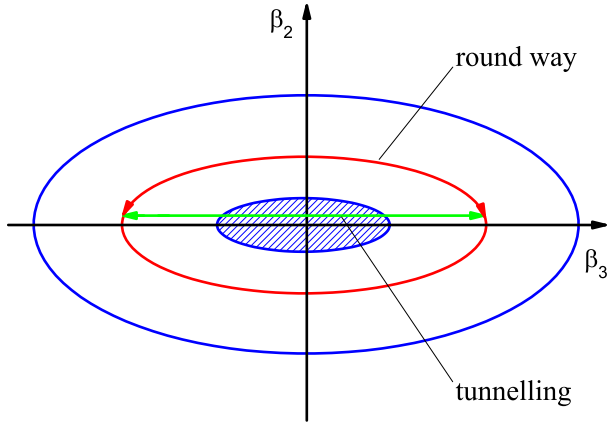}}} \caption{(Color
online) Schematic picture of the tunnelling and the rounding way
in the $\beta_3$- variable. See Sec. 2 for further discussion.}
\label{fig:04}
\end{figure}

In the case iii) of non-zero $\beta_{2min}$ and $\beta_{3min}$,
Eq.~(\ref{CL}) imposes the relation
\begin{eqnarray}
\frac{d_2}{C_2}=\frac{d_3}{C_3} \ . \label{dcratio}
\end{eqnarray}
It determines an elliptic form of the bottom of the
two-dimensional potential surface given by
$\beta_{2min}^2/\sqrt{2X(I)/(d_2C_2)}+
\beta_{3min}^2/\sqrt{2X(I)/(d_3C_3)} =1$. The shape of the
potential corresponding to the case iii) is illustrated in Fig. 3.
It suggests that the system moves in the two-dimensional space of
the deformation variables $\beta_2$ and $\beta_3$ by rounding the
internal potential core. If a {\em prolate quadrupole deformation} is
considered, the rounding is performed between positive and negative
$\beta_3$ values in the space of $\beta_2>0$.  This situation can
be considered as the two-dimensional extension of the one-dimensional
case in which the $\beta_2$ coordinate is frozen. To explain
this in detail, we consider a horizontal (equipotential) intersection
of the shape in Fig. 3, which is illustrated schematically in Fig. 4.
We see that if the quadrupole coordinate is fixed at some value of
$\beta_2>0$, the motion in the octupole coordinate between positive
and negative $\beta_3$-values is characterized by the tunnelling
through a potential barrier (a vertical intersection of the core).
When $\beta_2$ is let to vary, the tunnelling is replaced by a motion
along the curved way rounding the potential core.

The above case iii) is of particular interest, due to the
simultaneous presence of non-zero coordinates of the potential
minimum in both degrees of freedom. It suggests that the oscillations
in the quadrupole and octupole coordinates are involved in the collective
motion on the same footing. As it will be seen below, such a situation
appears to take place in certain nuclear regions. Moreover, it will be
seen that the ellipsoidal symmetry in the potential bottom allows, under
some additional conditions, a {\em complete analytic determination} of
the energy spectrum. This is why in the following we shall imply
this case, unless something different is indicated. Also, we
assume only the presence of prolate quadrupole deformations. This is
why hereafter we consider only the $\beta_2>0$ part of the space.

Further, we examine the evolution of the potential shape with the
angular momentum $I$. We consider the following two cases.

I) The potential minimum (the 2-dimensional bottom) is allowed to
change with $I$ for fixed values of the stiffness parameters $C_2$
and $C_3$.

II) The minimum is fixed, so that the values $\beta_{2min}$ and
$\beta_{3min}$ determine an ellipse which does not change with the
angular momentum.

It is clear that in the case I) of fixed stiffness parameters, the
quadrupole and the octupole deformations corresponding to the
potential minimum should exhibit an overall increase in the denominator of
Eqs.~(\ref{CL}) with increasing $I$.

In the case II) of fixed minima, the stiffness parameters
$C_2\equiv C_2(I)$ and $C_3\equiv C_3(I)$ increase quadratically
with $I$ according to the right hand sides of (\ref{CL}). Then the
substitution of Eqs.~(\ref{CL}) into (\ref{Ub2b3I}), leads to the
following form of the quadrupole--octupole potential
\begin{eqnarray}
U(\beta_2,\beta_3,I)
&=&X(I)\left(\frac{d_2\beta_{2}^2+d_3\beta_{3}^2}
{(d_2\beta_{2min}^2+d_3\beta_{3min}^2)^2}+
\frac{1}{d_2\beta_2^2+d_3\beta_3^2}\right)\ . \label{octfixL}
\end{eqnarray}
If the origin of the energy scale is fixed at the potential minimum,
one has
\begin{eqnarray}
U_{I}(\beta_2,\beta_3)&=&U(\beta_2,\beta_3,I)-
U(\beta_{2min},\beta_{3min},I) \nonumber \\
&=&\frac{X(I)[d_2(\beta_2^2-\beta_{2min}^2)+
d_3(\beta_3^2-\beta_{3min}^2)]^2}
{(d_2\beta_{2min}^2+d_3\beta_{3min}^2)^2(d_2\beta_2^2+d_3\beta_3^2)}
. \label{octfix}
\end{eqnarray}

We remark that the potential (\ref{octfixL}) includes the rotational
contribution of the centrifugal term, which moves up the energy with
increasing angular momentum $I$. On the other hand, in the potential
(\ref{octfix}) the explicit contribution of the rotational degree
of freedom is diminished, so that the energy term keeps mainly the
vibrational component. The shape of the potential (\ref{octfix}) with
$\beta_2 >0$ is illustrated in Fig. 5.

\begin{figure}[t]
\epsfxsize=10.cm   %or \epsfysize= HEIGHT cm
\centerline{\epsfbox{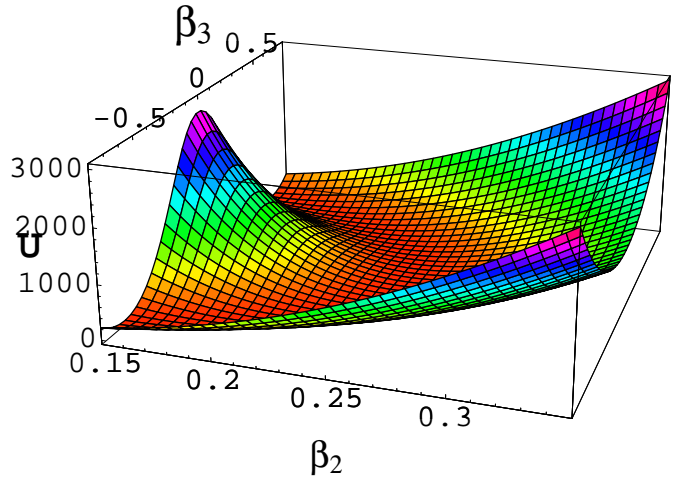}} \caption{(Color online) 3D plot of
the potential $U_I(\beta_2,\beta_3)$, Eq.~(\protect\ref{octfix}),
in MeV for $I=5$ as a function of $\beta_2$ and $\beta_3$ with
$\beta_{2min}=0.25$, $\beta_{3min}=0.1$, $d_2=0.1$
$\hbar^2$MeV$^{-1}$, and $d_3=0.01$ $\hbar^2$MeV$^{-1}$.}
\label{fig:05}
\end{figure}

\section{Model potentials and the Schr\"{o}dinger equation
in polar variables}

Further, it is convenient to introduce polar variables $\eta$ and
$\phi$ by taking
\begin{eqnarray}
\beta_2=\frac{\eta }{\sqrt{d_2/d}}\cos\phi \ ;\qquad
\beta_3=\frac{\eta }{\sqrt{d_3/d}}\sin\phi \ ,
\end{eqnarray}
with $d=(d_2+d_3)/2$. Considering $\beta_2 >0$, we have
\begin{eqnarray}
\eta=\frac{1}{\sqrt{d}}\sqrt{d_2\beta_2^2+d_3\beta_3^2}\ ; \qquad
\phi=\arctan\left( {\frac{\beta_3}{\beta_2}
\sqrt{\frac{d_3}{d_2}}}\right )\ , \label{phi}
\end{eqnarray}
where the ``effective'' deformation variable $\eta$ is defined with
positive values $\eta >0$, while the relative (``angular'') variable
$\phi$ is defined in the interval $-\pi /2\leq\phi\leq\pi /2$.
We remark that the negative $\phi$- values correspond to
negative $\beta_3$. (The variable $\beta_3$ takes both positive
and negative values.)

Then the quadrupole-octupole Hamiltonian (\ref{Hqo}) can be
written in the form
\begin{eqnarray}
H_{qo}&=&-\frac{\hbar^2\,d_2}{2dB_2}\left[ \cos^2{\phi}
\frac{\partial^2}{\partial \eta ^2}+\frac{1}{\eta }\sin^2{\phi}
\frac{\partial}{\partial \eta }+\frac{1}{\eta ^2}\sin^2{\phi}
\frac{\partial^2}{\partial \phi^2} \right. \nonumber \\
&+& \left. 2\frac{1}{\eta ^2}\sin{\phi}\,cos{\phi}\frac{\partial}
{\partial \phi}-2\frac{1}{\eta }\sin{\phi}\,cos{\phi}
\frac{\partial^2}{\partial \eta  \partial \phi}\right] \nonumber\\
&-&\frac{\hbar^2 \, d_3}{2dB_3}\left[ \sin^2{\phi}
\frac{\partial^2}{\partial \eta ^2}+\frac{1}{\eta }\cos^2{\phi}
\frac{\partial}{\partial \eta }+\frac{1}{\eta ^2}\cos^2{\phi}
\frac{\partial^2}{\partial \phi^2} \right. \nonumber \\
&-&\left. 2\frac{1}{\eta ^2}\sin{\phi}\,cos{\phi}\frac{\partial}
{\partial \phi}+2\frac{1}{\eta }\sin{\phi}\,cos{\phi}
\frac{\partial^2}{\partial \eta  \partial \phi}\right]+\nonumber\\
&+&U_I(\eta) \ . \label{Hqo01}
\end{eqnarray}

Under the condition (\ref{dcratio}), the potential energy depends
only on the effective deformation  variable $\eta$ and on the
angular momentum $I$, and not on the relative (angular) variable
$\phi$. Then in the case I) of fixed stiffness parameters one has
\begin{eqnarray}
U_{I}(\eta)=\frac{1}{2}C\eta^2+\frac{X(I)}{d\eta^2}\ \qquad\qquad
(\mathrm{ Case\ I}), \label{cpotenc}
\end{eqnarray}
where $C$ is defined according to Eq. (\ref{dcratio}) as
$1/C=d_2/(dC_2)=d_3/dC_3$.

In the case II) of fixed minima the potential term appears in the
following two forms
\begin{eqnarray}
U_{I}(\eta )&=& X(I)\frac{\eta ^4+\eta_{min}^4} {d\eta_{min}^4 \,
\eta ^2} \ \ \ \qquad\qquad (\mathrm{ Case\ II.A});
\label{octfix00}\\
U_{I}(\eta )&=& X(I)\frac{[\eta ^2-\eta_{min}^2]^2} {d\eta_{min}^4
\, \eta ^2} \qquad\qquad (\mathrm{ Case\ II.B}), \label{octfix01}
\end{eqnarray}
where Eq.~(\ref{octfix00}) corresponds to the rotation dependent
potential (\ref{octfixL}), while Eq.~(\ref{octfix01}) represents
the essentially vibrational term (\ref{octfix}). The quantity
$\eta_{min}=(1/\sqrt{d})(d_2\beta_{2min}^2+d_3\beta_{3min}^2)^{1/2}$
is the value of the variable $\eta$ in the potential minimum. In
the following we shall refer to  Eq.~(\ref{octfix00}) as case II.A
and to Eq.~(\ref{octfix01}) as case II.B.

Using the effective deformation variable $\eta$ we are also capable
to examine a third case (III) of an infinite square well with an
infinite core at zero defined as
\begin{eqnarray}
U^{I}_{\mathrm{w}}(\eta )&=& \left\{ \begin{array}{cc}
\frac{X(I)}{d\eta^2}  & \eta \leq \eta_{\mathrm{w}} \\
\infty & \ \ \eta > \eta_{\mathrm{w}}  \
\end{array}\right. \qquad\qquad (\mathrm{ Case\ III}),
\label{sqwell}
\end{eqnarray}
where $\eta_{\mathrm{w}}$ is a parameter determining the width of
the well.

Now we assume the following relation between the quadrupole and
octupole mass and inertia parameters
\begin{eqnarray}
\frac{d_2}{dB_2}=\frac{d_3}{dB_3}=\frac{1}{B} \ .
\label{dbratio}
\end{eqnarray}
This leads to the following form of the model Hamiltonian
\begin{eqnarray}
H_{qo}&=&-\frac{\hbar^2}{2B}\left[\frac{\partial^2}{\partial\eta^2}+
\frac{1}{\eta }\frac{\partial}{\partial \eta }+
\frac{1}{\eta ^2}\frac{\partial^2}{\partial \phi^2} \right]
+U_I(\eta ) \ .
\label{Hqo02}
\end{eqnarray}
The assumption (\ref{dbratio}), which much simplifies the problem,
suggests that  $d_2$ and $d_3$ are related to the mass parameters
$B_2$ and $B_3$ respectively,  through the same coefficient $d/B$.
By comparing Eq.~(\ref{dbratio}) and Eq.~(\ref{dcratio}), we obtain
$C_2/B_2=C_3/B_3$, or $\omega_2^2=\omega_3^2$, i.e. the assumption
(\ref{dbratio}) implies that both degrees of freedom, quadrupole
and octupole, are characterized by equal angular frequencies
$\omega_2$ and $\omega_3$, respectively. This means that a
{\em coherent interplay} between the two collective modes is assumed.
In other words, the condition  (\ref{dbratio}) suggests that the
oscillations in the quadrupole and octupole coordinates are represented
in the collective motion on the same footing. The quantity $B$ in
Eq.~(\ref{dbratio}) has the meaning of the effective mass of the
total quadrupole--octupole system.

The Schr\"{o}dinger equation for the Hamiltonian (\ref{Hqo02}) has
the form
\begin{eqnarray}
-\frac{\hbar^2}{2B}\frac{1}{\eta ^2} \left[ \frac{\eta
^2\partial^2}{\partial \eta ^2}+ \eta \frac{\partial}{\partial
\eta }+ \frac{\partial^2}{\partial \phi^2} \right]\Phi (\eta
,\phi) +U_I(\eta )\Phi (\eta ,\phi)=E\Phi (\eta ,\phi) \ .
\label{ScrHqo}
\end{eqnarray}
After dividing it by $\hbar^2/(2B\eta ^2)$ and separating the
variables $\eta $ and $\phi$ through $\Phi(\eta ,\phi)=\psi(\eta )
\varphi(\phi)$ we obtain the following two equations
\begin{eqnarray}
\frac{\partial^2}{\partial \eta ^2}\psi(\eta )+ \frac{1}{\eta
}\frac{\partial}{\partial \eta }\psi(\eta )
+\frac{2B}{\hbar^2}\left[ E-\frac{\hbar^2} {2B}\frac{k^2}{\eta
^2}-U_I(\eta )\right] \psi (\eta )&=&0 \ ; \label{Hqo03} \\
\frac{\partial^2}{\partial
\phi^2}\varphi(\phi)+k^2\varphi(\phi)&=&0 \ , \label{wphi}
\end{eqnarray}
where $k$ is the separation quantum number.

\section{Analytic solutions and numerical results}

In the following we give analytic solutions of the above equations
in the cases I-III with the potentials (\ref{cpotenc}),
(\ref{octfix00}), (\ref{octfix01}), and (\ref{sqwell}).

In the {\bf Case I}, after introducing the potential
(\ref{cpotenc}) into Eq.~(\ref{Hqo03}) we have
\begin{eqnarray}
\frac{\partial^2}{\partial \eta ^2}\psi(\eta )+ \frac{1}{\eta
}\frac{\partial}{\partial \eta }\psi(\eta )
+\frac{2B}{\hbar^2}\left[E-\frac{\hbar^2} {2B}\frac{k^2}{\eta ^2}-
\frac{1}{2}C\eta ^2 - \frac{X(I)}{d\eta ^2} \right] \psi (\eta )=0
\ . \label{Hqo004}
\end{eqnarray}
By introducing a reduced energy $\varepsilon =
\frac{2B}{\hbar^2}E$ and a reduced angular momentum factor
$\widetilde{X}(I) =bX(I)$, with $b=\frac{2B}{\hbar^2 d}$, we
obtain Eq.~(\ref{Hqo004}) in the form
\begin{eqnarray}
\frac{\partial^2}{\partial \eta ^2}\psi(\eta )+ \frac{1}{\eta}
\frac{\partial}{\partial \eta }\psi(\eta ) +\left[\varepsilon
-\frac{k^2+\widetilde{X}(I)}{\eta ^2}- \frac{BC}{\hbar^{2}}\eta ^2
 \right] \psi (\eta )=0 \ .
\label{Hqo005}
\end{eqnarray}
The effective potential appearing in the brackets of
Eq.~(\ref{Hqo005}) is of a form similar to the Davidson potential
\cite{Dav32}, which is analytically solvable \cite{Elliott,Rowe}.
Thus Eq.~(\ref{Hqo005}) can be solved analytically and we
obtain the following explicit expression for the energy spectrum
\begin{equation}
E_{n,k}(I) =\hbar\omega \left[ 2n+1+\sqrt{k^2+\widetilde
X(I)}\right], \label{spect1a}
\end{equation}
where $\omega=\sqrt{C/B}$ and $n=0,1,2,...$. The eigenfunctions
$\psi(\eta )$ of Eq.~(\ref{Hqo004}) are obtained in terms of the
Laguerre polynomials
\begin{equation}
\psi^I_n(\eta )=\sqrt {\frac {2\Gamma(n+1)}{\Gamma(n+2s+1)}}
e^{-a\eta^2/2}a^s\eta^{2s}L^{2s}_n(a\eta^2)\ , \label{psieta1}
\end{equation}
where $a=\sqrt{BC}/\hbar$ and $s=\sqrt{k^2+\widetilde{X}(I)}/2$.

Now we remark that Eq.~(\ref{wphi}) in the variable $\phi$ is
solved under the periodic boundary condition $\varphi(\phi
+2\pi)=\varphi(\phi)$. On the other hand the assumption $\beta_2 >0$,
which is equivalent to the consideration of an infinite potential wall
at $\beta_2 =0$ (or $\phi=\pm\pi /2$), imposes the additional condition
\begin{eqnarray}
\varphi(-\pi /2)=\varphi(\pi /2)=0\ . \label{pib}
\end{eqnarray}
Eq.~(\ref{wphi}) has two different solutions
satisfying the condition (\ref{pib}) with positive, $\pi =(+)$,
and negative, $\pi =(-)$, parity as follows
\begin{eqnarray}
\varphi^{+}(\phi)&=& \sqrt{2/\pi}\cos (k\phi ) \ , \qquad k=\pm 1,
\pm 3, \pm 5,
...\ ;\label{parplus} \\
\varphi^{-}(\phi)&=& \sqrt{2/\pi}\sin (k\phi ) \ , \qquad k=\pm 2,
\pm 4, \pm 6, ...\ . \label{parminus}
\end{eqnarray}

Eq.~(\ref{parplus}) provides positive parity for the intrinsic
wave function, while Eq.~(\ref{parminus}) corresponds to a
negative parity function. As a result, the intrinsic wave function
appears in the form $\Phi^{\pm}(\eta ,\phi)=\psi(\eta )
\varphi^{\pm}(\phi)$. On the other hand, the
$\mathcal{R}\mathcal{P}$- symmetry of the total wave function of
the system, $\Psi\sim \Phi^{\pm}(\eta ,\phi)|IKM\rangle$, has to be
conserved. The $\mathcal{R}$- symmetry of the rotation function
$|IKM\rangle $ is characterized by the factor $(-1)^I$. For the
total state of the system one has $\pi (-1)^I=1$. It follows
that for $I$=even the quantum number $k$ is allowed to take the
values $k_{(+)}=1,3,5,\dots$, corresponding to the even function
(\ref{parplus}), while for $I$=odd one has $k_{(-)}=2,4,6, \dots$
corresponding to the odd function (\ref{parminus}). Thus, when the
angular momentum is changed from $I$=odd to $I$=even and vice
versa, the respective values of the quantum number $k$ should
switch between $k_{(-)}$ and $k_{(+)}$. This parity effect
provides an odd-even staggering structure of the spectrum
(\ref{spect1a}). We consider that the lowest states of the system
with respect to the variable $\phi$ are characterized by
the lowest $k$- values, $k_{(+)}=1$ for $I=$even and
$k_{(-)}=2$ for $I=$odd. Therefore, the staggering behavior of
the model spectrum is provided by the difference $\Delta
k^{2}=k_{(-)}^{2}-k_{(+)}^{2}=3$.

\begin{figure}[t]
\centerline{{\epsfxsize=6.5cm\epsfbox{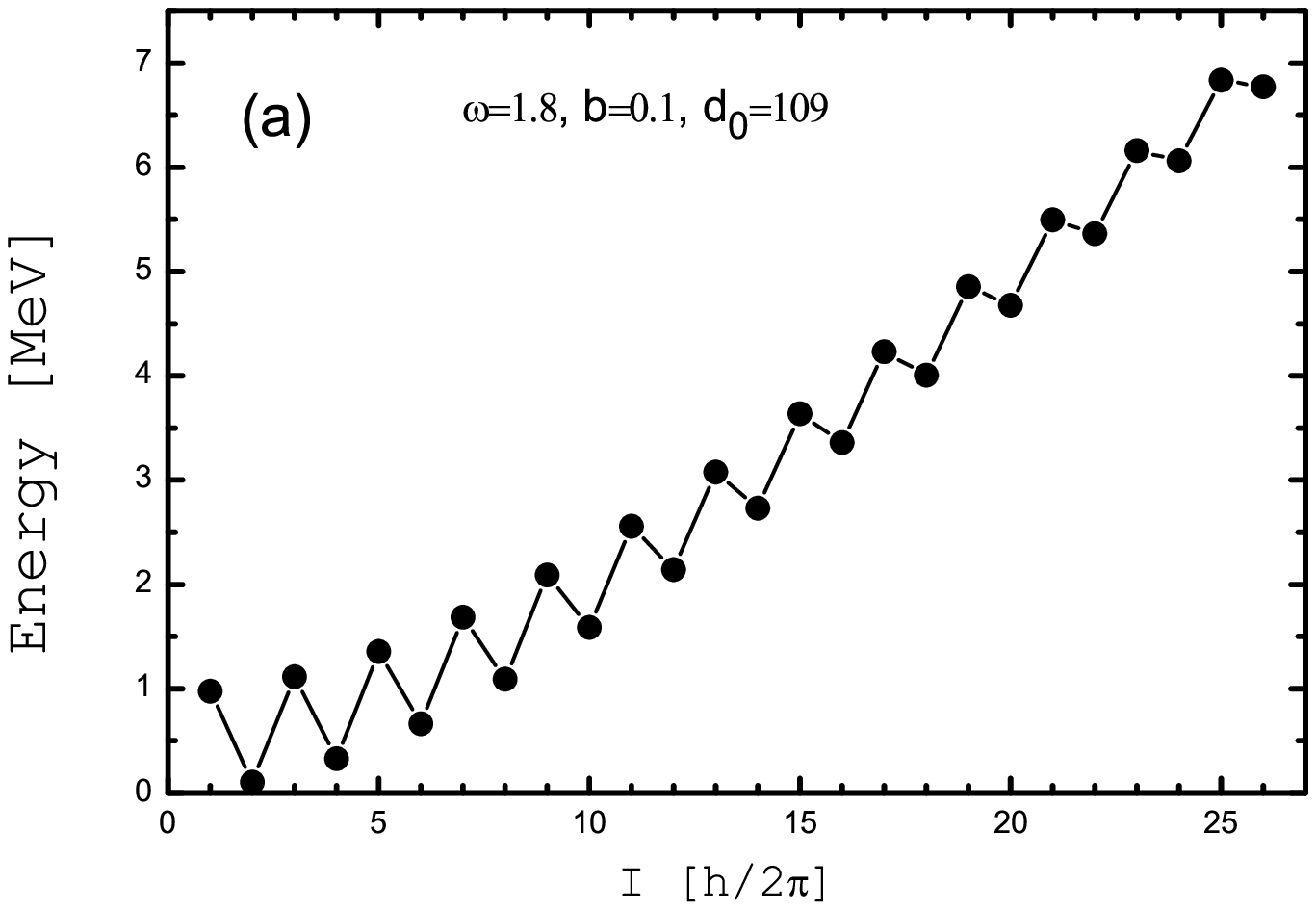}}\ \
{\epsfxsize=6.5cm\epsfbox{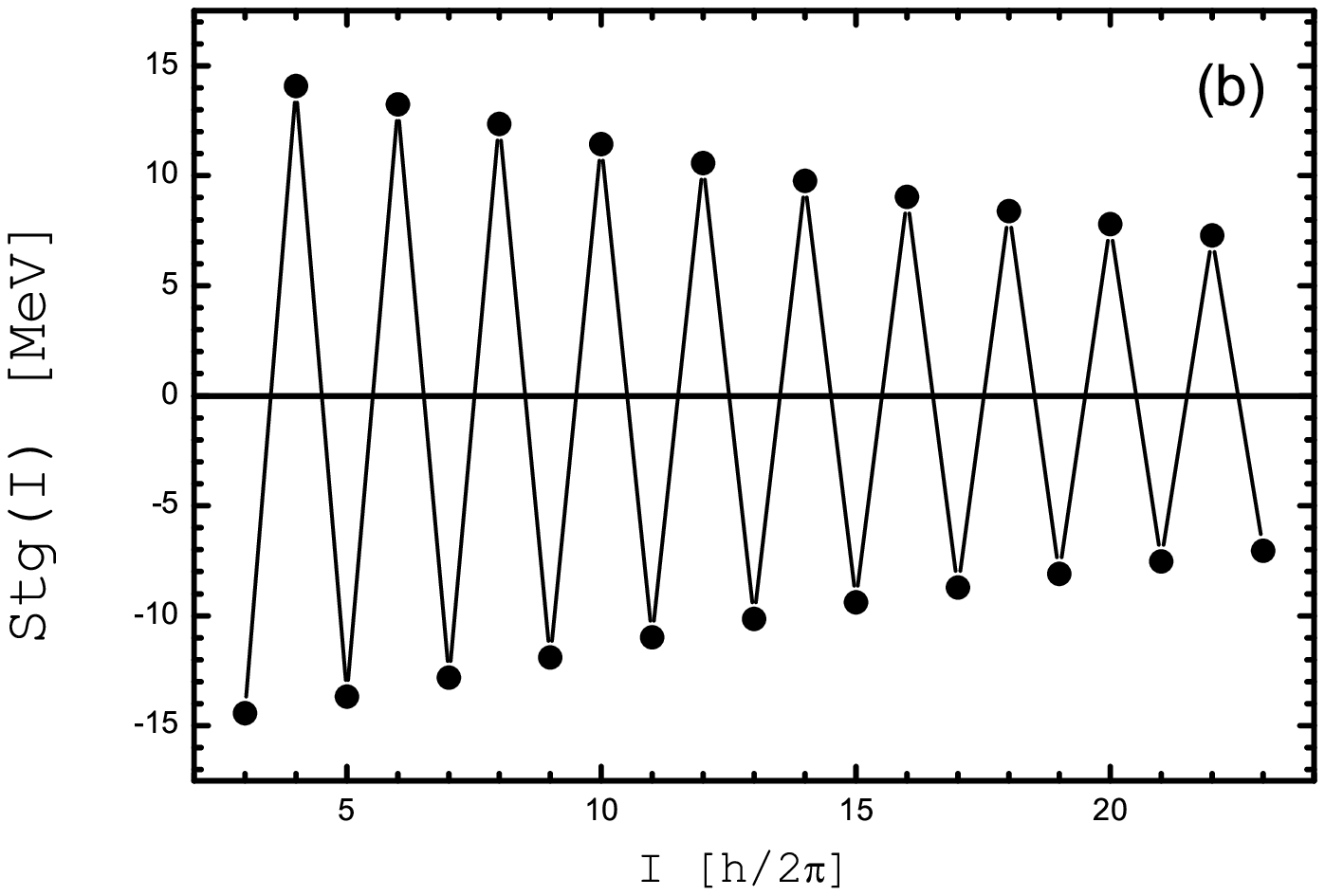}}} \caption{Schematic energy
levels (a) and staggering pattern (b) for the spectrum
(\protect\ref{spect1a}) with $n=0$. The parameter values are shown
in part (a), $\omega$ is given in MeV$/\hbar$, while $b$ and $d_0$
are in $\hbar^{-2}$ and $\hbar^2$ respectively.} \label{fig:06}
\end{figure}

In such a way the energy expression (\ref{spect1a}), with the
parity-dependent quantum number $k$, determines the structure
of an alternating parity spectrum. The energy levels $E_{0,k}(I)$,
with $n=0$, correspond to the yrast alternating parity sequence.
The levels with $n\neq0$ correspond to higher energy bands, in which
the rotational states are built on quadrupole--octupole (mixed
$\beta_2$-$\beta_3$) vibrations of the system.
In this case, the states with even $I$ appear similarly to the states
of a higher $\beta$- (quadrupole) band. Thus, the present model suggests
that, in the nuclear regions with quadrupole--octupole collectivity, one
may consider ``octupole mixed'' $\beta$- band structures. We remark
that, in the present model framework, the $\gamma$- bands are not included.
This can be done in an extended formalism allowing the simultaneous
consideration of the $\gamma$- variable. In addition, the octupole
triaxiality can be taken into account. Then one may also discuss
possible ``octupole admixtures'' in the $\gamma$- band structure.

\begin{figure}[t]
\centerline{\epsfxsize=14.cm\epsfbox{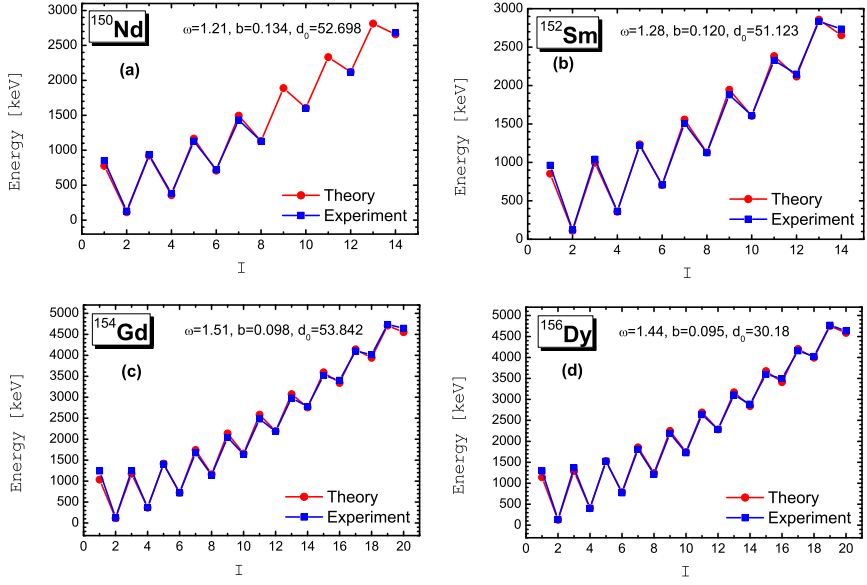}} \caption{(Color
online) Theoretical and experimental energy levels for the
alternating parity bands in $^{150}$Nd (data from \cite{Nd150en}),
$^{152}$Sm (data from \cite{Sm152en}), $^{154}$Gd (data from
\cite{Gd154en}) and $^{156}$Dy (data from \cite{Dy156en}). The
theoretical results are obtained by (\protect\ref{spect1a}) with
$n=0$. The parameter units are as in Fig. 6.} \label{fig:07}
\end{figure}

\begin{figure}[t]
\centerline{\epsfxsize=14.cm\epsfbox{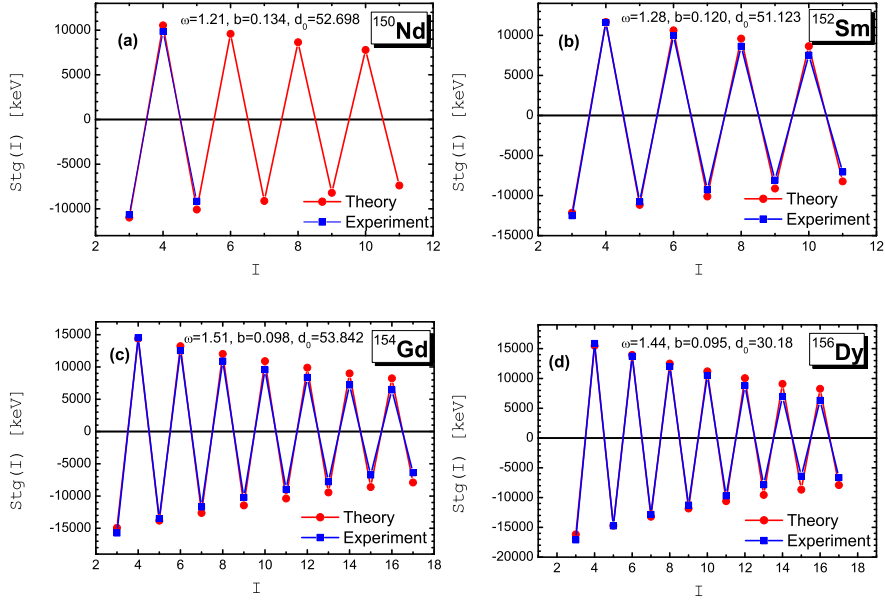}} \caption{(Color
online) Theoretical and experimental staggering patterns for the
alternating parity bands in $^{150}$Nd, $^{152}$Sm, $^{154}$Gd and
$^{156}$Dy. The theoretical results are obtained by
(\protect\ref{spect1a}) with $n=0$. The parameter units are as in
Fig. 6.} \label{fig:08}
\end{figure}

We can estimate analytically the staggering effect at higher
angular momenta, where $\Delta k^{2}\ll \widetilde X(I)$. The
square root term $\sqrt{k^2+\widetilde X(I)}$ in
Eq.~(\ref{spect1a}) can be expanded as $\sqrt{\widetilde
X(I)}+k^2/(2\sqrt{\widetilde X(I)})$. We see that the term
$k^2/(2\sqrt{\widetilde X(I)})$, which is responsible for the
staggering effect, decreases nearly linearly with the angular
momentum $I$ since $\widetilde X(I)=bX(I)\sim I(I+1)$. We consider
the quantity $b=2B/(\hbar^2d)$ as a model parameter. The numerical
behavior of the energy and the staggering effect for
the spectrum (\ref{spect1a}) is illustrated in Fig.~6. The
staggering effect is illustrated in terms of the five point
quantity
\begin{equation}
\mbox{Stg}(I)= 6\Delta E(I)-4\Delta E(I-1)-4\Delta E(I+1)+ \Delta
E(I+2)+\Delta E(I-2)\ , \label{stag}
\end{equation}
where $\Delta E(I)=E(I+1)-E(I)$. The schematic staggering pattern
suggests that the odd and even angular momentum sequences
approach each other towards higher angular momenta. It outlines a
trend for the forming of an octupole band. However, the linear
decrement of the staggering amplitude is not enough to provide such a
band structure at reasonable (experimentally observed) angular momenta.
A similar situation is observed in rare earth nuclei, where the alternating
parity levels approach each other without merging into a single band.

On this basis, we applied Eq.~(\ref{spect1a}) to describe the
alternating parity spectra in the nuclei $^{150}$Nd, $^{152}$Sm, $^{154}$Gd,
and $^{156}$Dy. The theoretical energies are obtained by taking
$\widetilde{E}_{n,k}(I)= E_{n,k}(I)-E_{n,k}(0)$, with $n=0$ and
$X(I)=\frac{1}{2}[d_0+I(I+1)]$, where the parameter $d_0$
characterizes the potential shape in the ground state, as mentioned
in the paragraph after Eq.~(\ref{Ub2b3I}). The parameters $\omega$,
$b$ and $d_0$ are adjusted to the energy levels by means of a least
square minimization procedure.

In Fig. 7 results for the energy levels of $^{150}$Nd,
$^{152}$Sm, $^{154}$Gd and $^{156}$Dy are compared to the
experimental data. The respective theoretical and experimental
staggering patterns are compared in Fig. 8. In $^{150}$Nd
[Fig. 7(a)], the levels with $I=9,11,13$ are predicted. The
respective staggering pattern for $I>5$ [Fig. 8(a)] is also
predicted. We see that in the nuclei $^{152}$Sm, $^{154}$Gd, and
$^{156}$Dy the experimental patterns confirm the predicted behavior
of alternating parity levels with increasing angular momentum.

In the {\bf Case II.A}, after introducing the potential
(\ref{octfix00}) into Eq.~(\ref{Hqo03}), we have the equation
\begin{eqnarray}
\frac{\partial^2}{\partial \eta ^2}\psi(\eta )+ \frac{1}{\eta
}\frac{\partial}{\partial \eta }\psi(\eta )
+\frac{2B}{\hbar^2}\left[E-\frac{\hbar^2} {2B}\frac{k^2}{\eta
^2}-\frac{X(I)}{d\eta ^2}- \frac{X(I)}{d\eta_{min}^4}\eta ^2
\right] \psi (\eta )=0 \ , \label{Hqo04}
\end{eqnarray}
which is solved in the same way as Eq.~(\ref{Hqo004}) of case I,
and the respective energy levels are obtained in the form
\begin{equation}
E_{n,k}(I)=\hbar^{2}\frac{\sqrt{\widetilde
X(I)}}{B\eta^2_{min}}\left[2n+1+\sqrt{k^2+\widetilde X(I)}\right],
\label{spect02a}
\end{equation}
where $n=0,1,2,\dots $. The eigenfunctions $\psi(\eta )$ of
Eq.~(\ref{Hqo04}) are of the same form as Eq.~(\ref{psieta1}) in
case I, but with $a=\sqrt{\widetilde X(I)}/\eta^2_{min}$.

\begin{figure}[t]
\centerline{{\epsfxsize=6.5cm\epsfbox{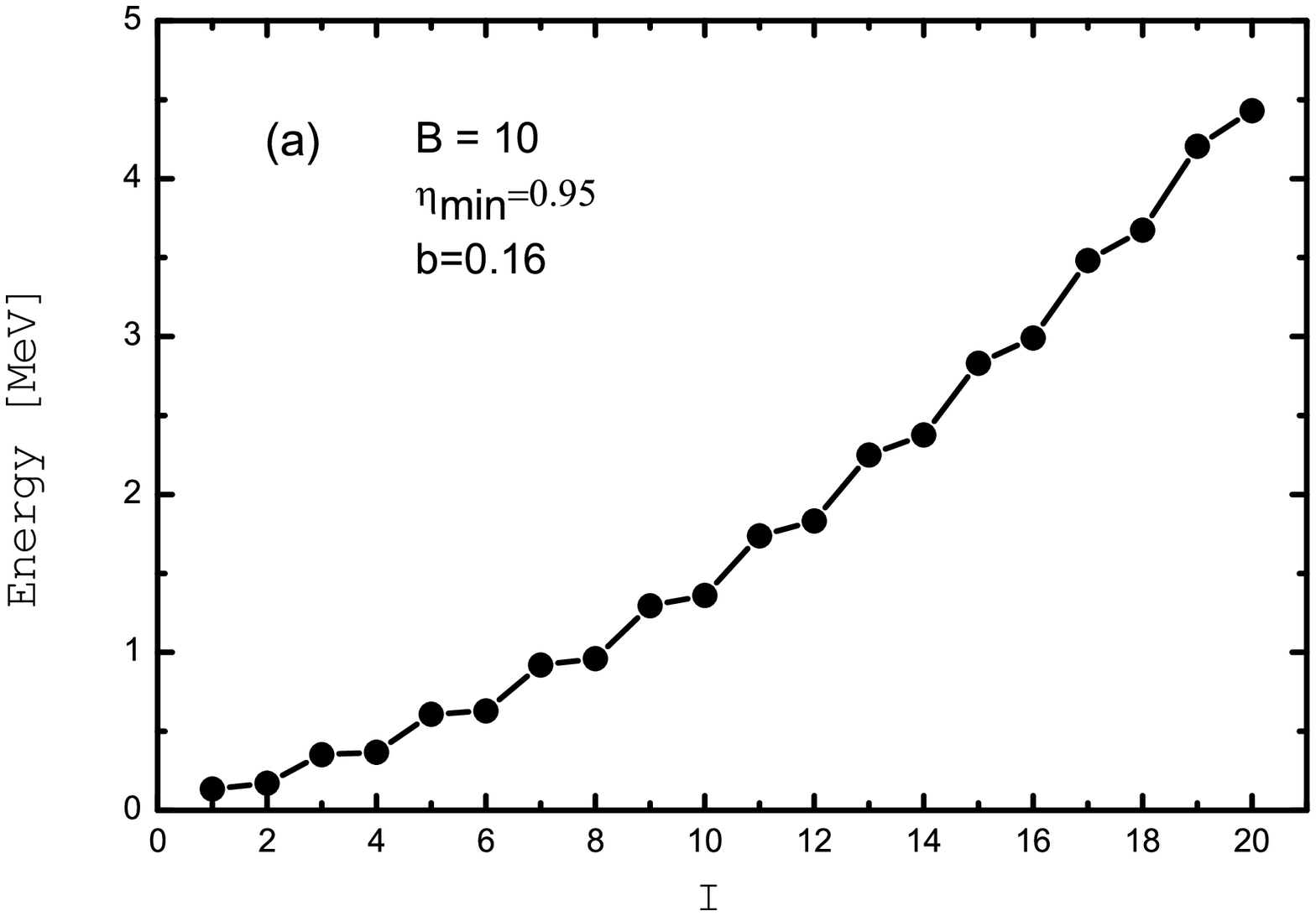}}\ \
{\epsfxsize=6.5cm\epsfbox{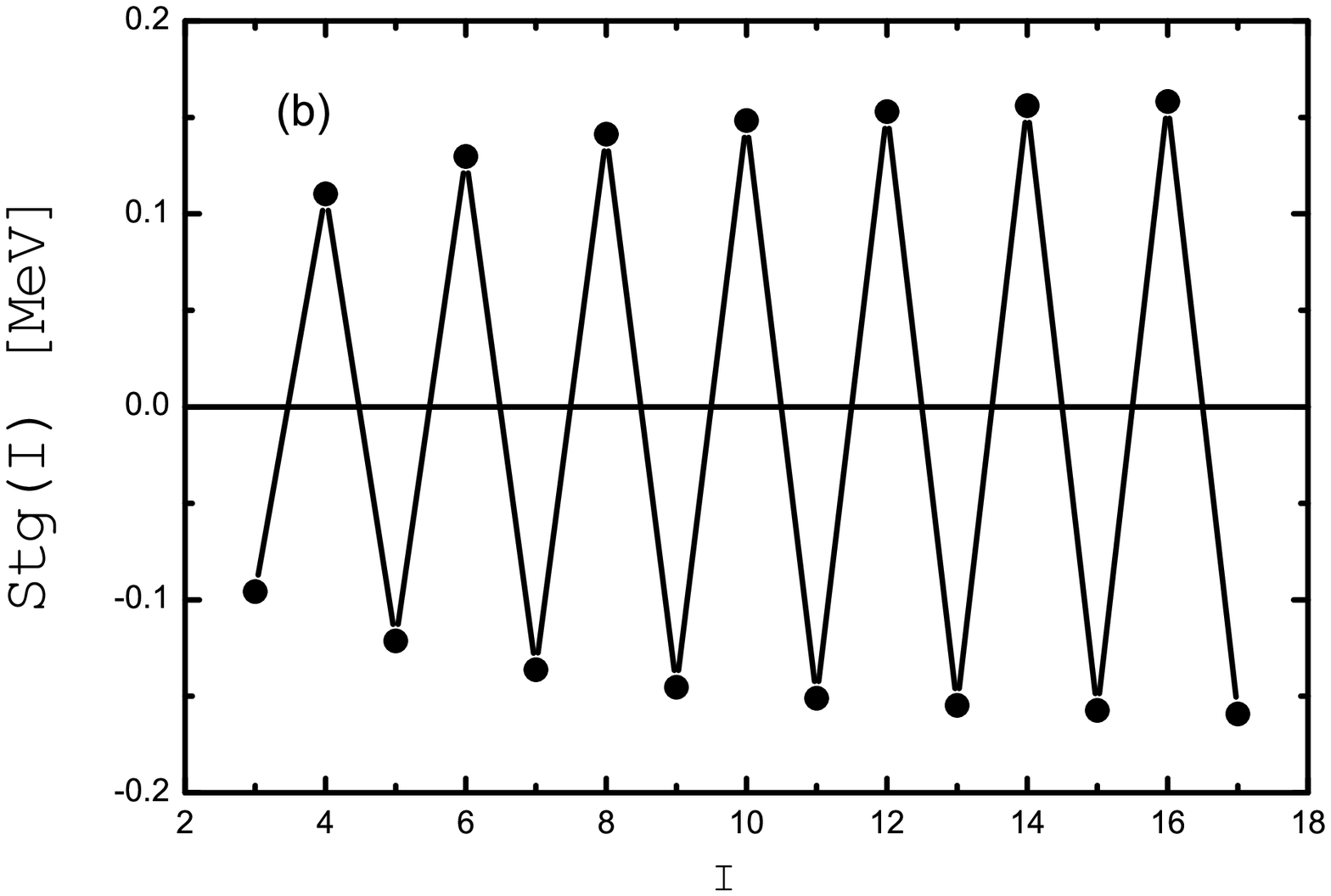}}} \caption{Schematic energy
levels (a) and staggering pattern (b) for the spectrum of
Eq.~(\protect\ref{spect02a}) with $n=0$. The parameter values are
shown in part (a), $B$ is given in $\hbar^2$/MeV, $b$ in
$\hbar^{-2}$, while $\eta_{min}$ is dimensionless.} \label{fig:09}
\end{figure}

\begin{figure}[t]
\centerline{{\epsfxsize=6.5cm\epsfbox{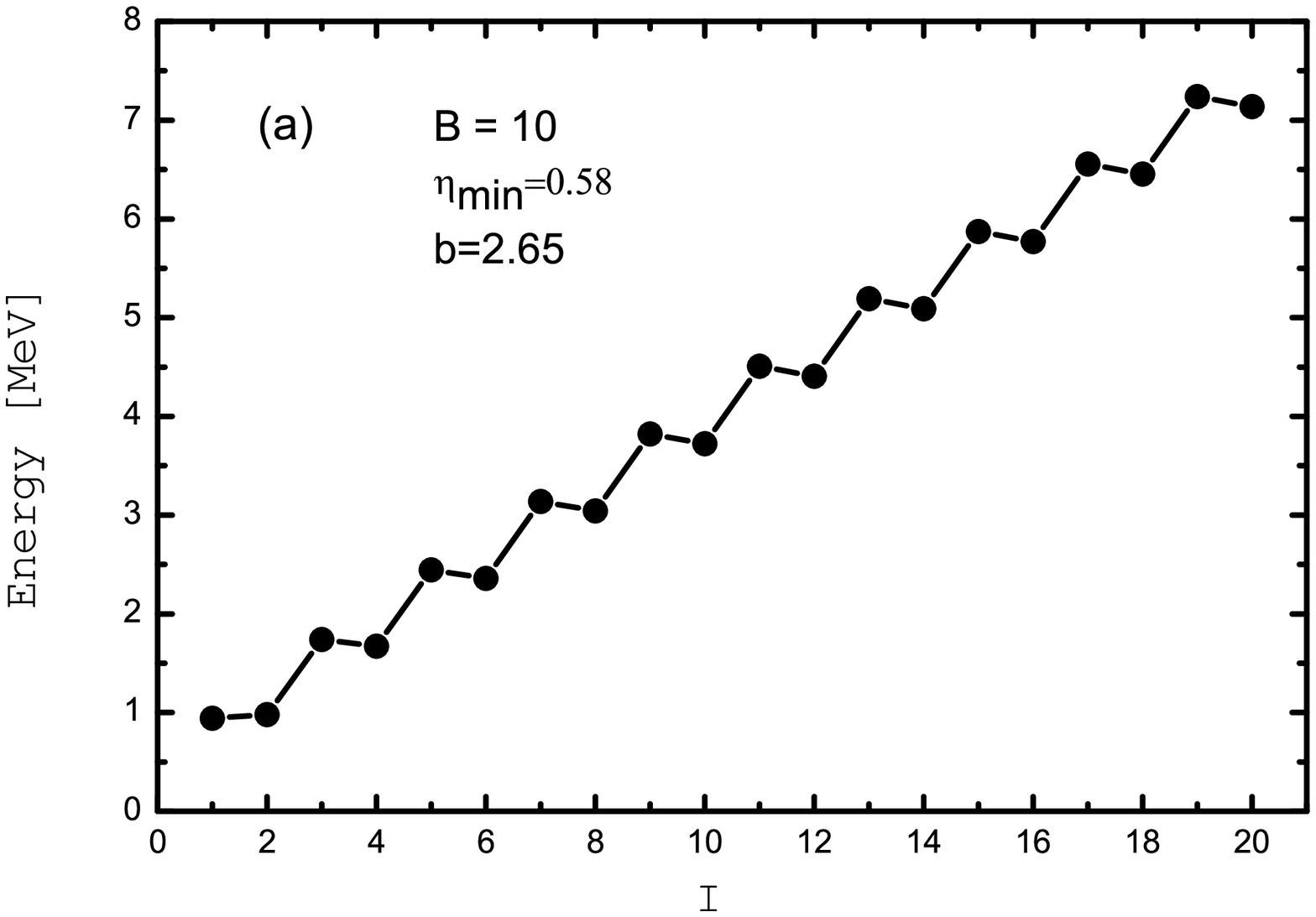}}\ \
{\epsfxsize=6.5cm\epsfbox{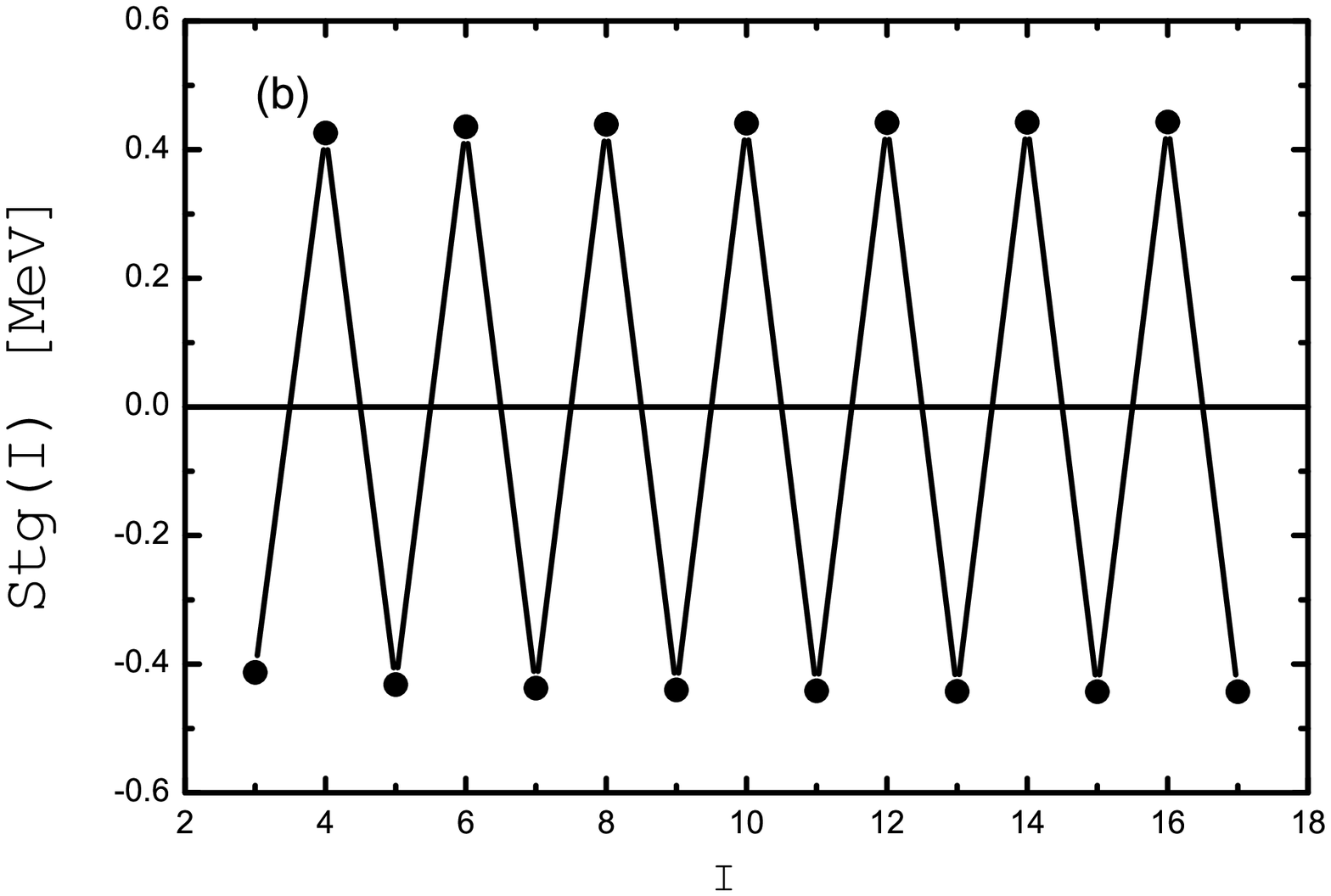}}} \caption{The same as in
Fig.~9, but for the spectrum of Eq.~(\protect\ref{spect02ab}).}
\label{fig:10}
\end{figure}

All considerations related to the $\phi$- equation (\ref{wphi})
and the quantum number $k$ are the same as in case I. However, now
we obtain a different behavior of the staggering amplitude as a
function of the angular momentum. This is seen after expanding the
term $\sqrt{\widetilde X(I)}\sqrt{k^2+\widetilde X(I)}$ of
Eq.~(\ref{spect02a}) in the form $\widetilde X(I)+k^2/2$. The
appearance of the staggering effect is only due to the term
$k^2/2$. Since the difference $\Delta k^{2}=3$ does not depend on
$I$, the staggering effect will be characterized by a {\em
constant} amplitude. The schematic behavior of the energy levels
and the respective staggering pattern for the spectrum of
Eq.~(\ref{spect02a}) are illustrated in Fig.~9. Indeed, we see
from Fig.~9(b) that, after some slight increase in the beginning,
towards the higher angular momenta the staggering amplitude
saturates to a constant value.

In {\bf Case II.B}, the potential (\ref{octfix01}) differs from
the potential (\ref{octfix00}) of case II.A by the term
$-2X(I)/(d\eta_{min}^2)$. The respective energy spectrum is
\begin{equation}
E_{n,k}(I)=\hbar^{2}\frac{\sqrt{\widetilde
X(I)}}{B\eta^2_{min}}\left[2n+1+\sqrt{k^2+\widetilde X(I)}
-\sqrt{\widetilde X(I)} \right ], \label{spect02ab}
\end{equation}
with the wave function $\psi(\eta )$ being the same as in case
II.A. The expression (\ref{spect02ab}) differs from
Eq.~(\ref{spect02a}) by the term $-\sqrt{\widetilde X(I)}$ in the
brackets. This term reduces the angular momentum dependence of the
energy to a linear (vibrational) behavior. On the other hand, it
does not affect the staggering effect. Therefore, similarly to the
case II.A, the staggering pattern for the levels (\ref{spect02ab})
will be characterized by a constant amplitude. This is seen from
the schematic numerical results illustrated in Fig.~10.

\begin{figure}[t]
\centerline{{\epsfxsize=6.5cm\epsfbox{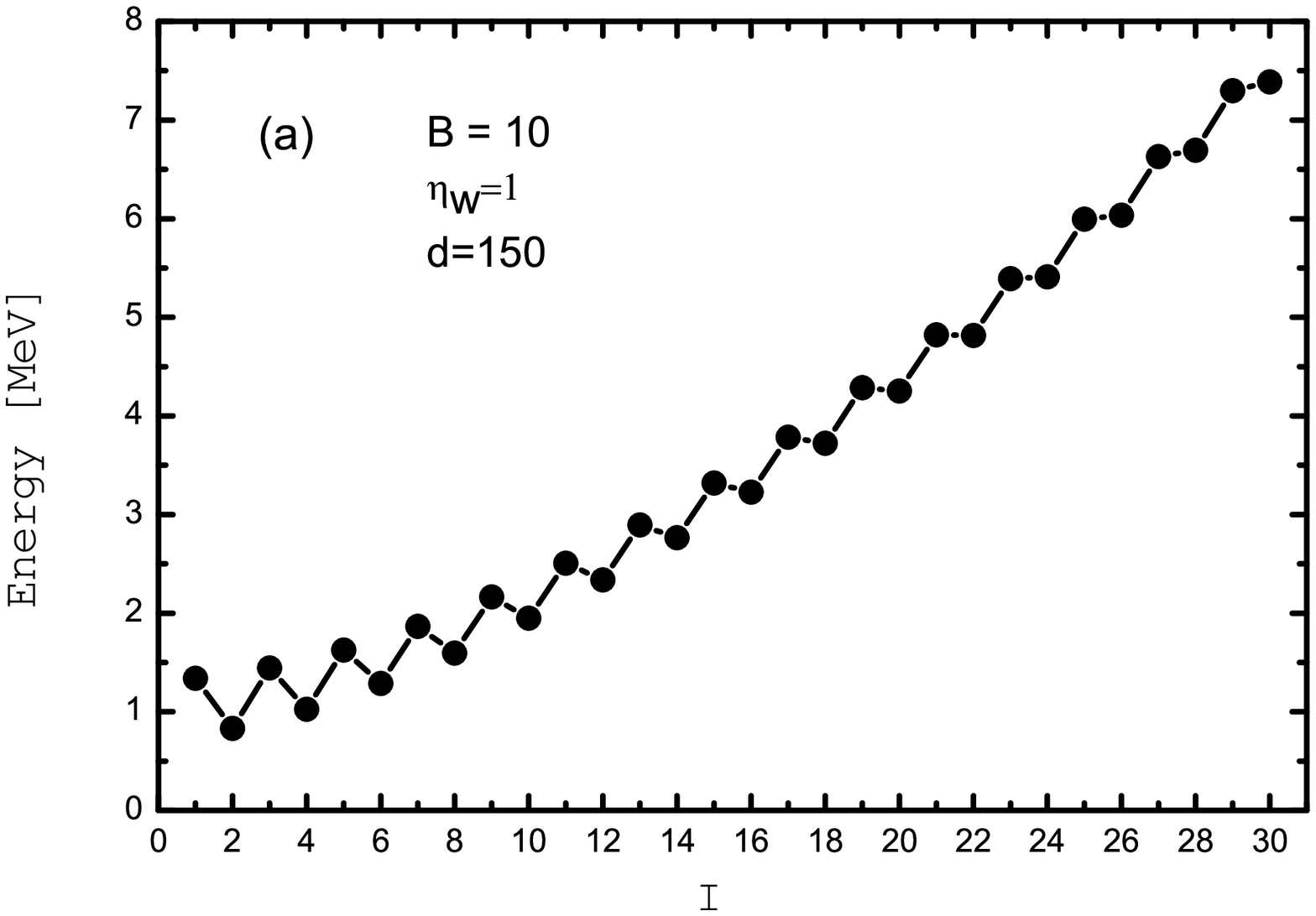}}\ \
{\epsfxsize=6.5cm\epsfbox{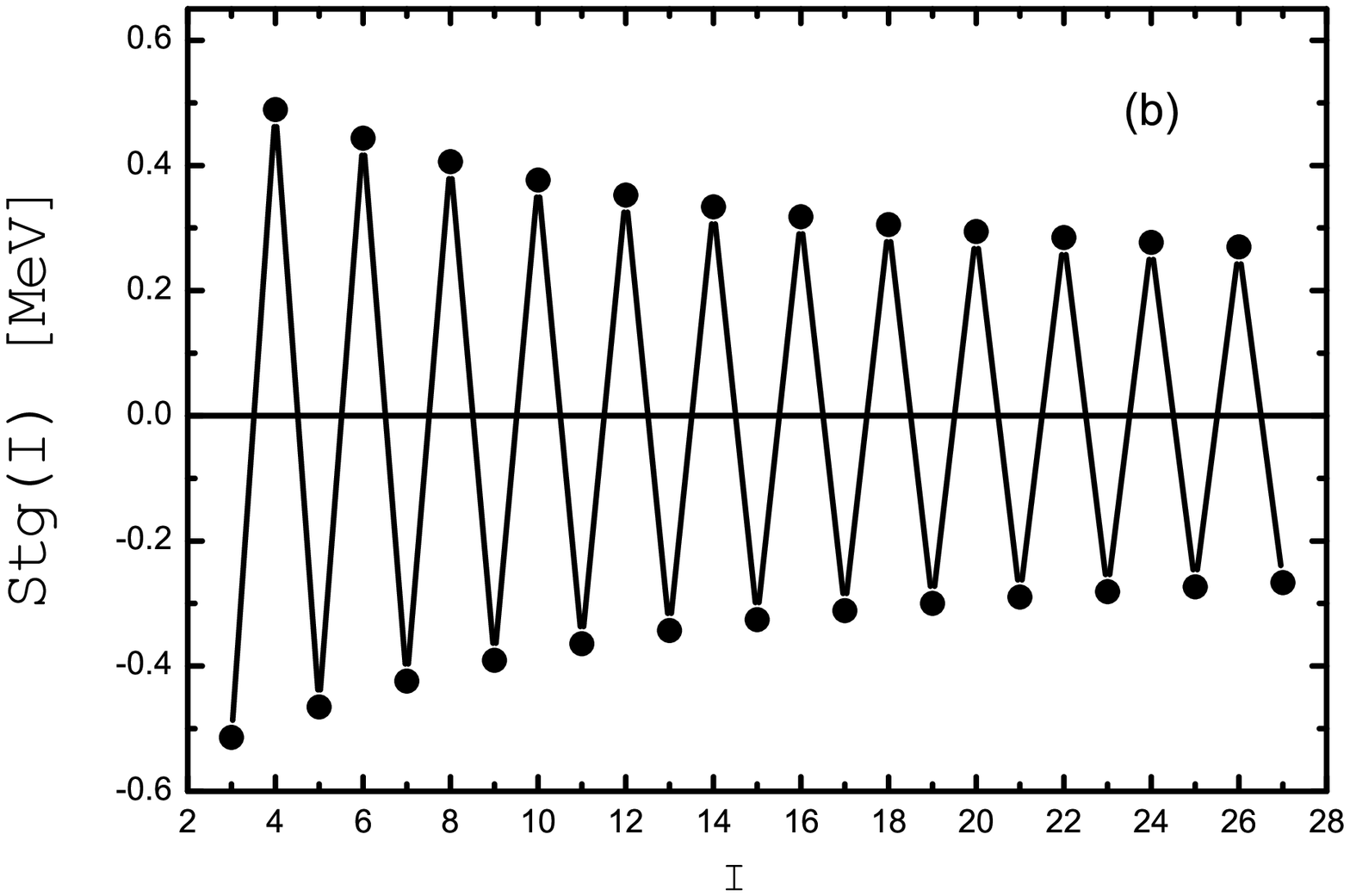}}} \caption{Schematic energy
levels (a) and staggering pattern (b) for the spectrum of
Eq.~(\protect\ref{sqspec}) with $n=0$. The parameter values are
shown in part (a), $B$ is given in $\hbar^2$/MeV, $d$ is in
$\hbar^2$MeV$^{-1}$, while $\eta_{\mathrm{w}}$ is dimensionless.}
\label{fig:11}
\end{figure}

In {\bf Case III} [the square potential well (\ref{sqwell})] the
Schr\"{o}dinger equation can be written in the form
\begin{equation}
\frac{\partial^2}{\partial\eta^2}\psi(\eta)+\frac{1}{\eta}
\frac{\partial}{\partial\eta}\psi(\eta)
+\left[\varepsilon-\frac{\nu^2}{\eta^2}\right]\psi(\eta)=0,
\label{sqeq}
\end{equation}
where $\nu^2=k^2+\widetilde{X}(I)$ and $\eta\leq
\eta_{\mathrm{w}}$ . By introducing new variables through the
definitions $z=\eta \kappa$ and $\varepsilon=\kappa^2$, we obtain
Eq.~(\ref{sqeq}) in the form of the Bessel equation
\begin{equation}
\frac{\partial^2}{\partial
z^2}\psi(z)+\frac{1}{z}\frac{\partial}{\partial z}\psi(z)
+\left[1-\frac{\nu^2}{z^2}\right]\psi(z)=0.
\label{besseq}
\end{equation}
The spectrum of this equation is determined by the boundary
condition $\psi_{\nu}(\eta_{\mathrm{w}})=0 $, and is given by
\begin{equation}
\varepsilon=\kappa_{\nu,\,n}^2,
\qquad\kappa_{\nu,\,n}=\frac{x_{\nu,\,n}}{\eta_{\mathrm{w}}}\ ,
\label{sqspec}
\end{equation}
where $x_{\nu,\,n}$ is  the $n$-th  zero of the Bessel function
$J_{\nu}(z)$. The eigenfunctions have the form
$\psi_{\nu,\,n}(\eta)=c_{\nu,\,n}J_{\nu}(\kappa_{\nu,\,n}\eta)$,
where $c_{\nu,\,n}$ are normalization constants. The schematic
behavior of the spectrum (\ref{sqspec}) and the respective
staggering pattern are illustrated in Fig.~11. We remark that the
staggering amplitude initially decreases, while towards higher
$I$ it saturates to a constant value.

\section{Electric transition probabilities}

The formalism developed so far allows the calculation of E1, E2
and E3 transition probabilities for the energy spectra in the
considered cases I-III.  In the cases I and II, the reduced
probability for an electric transition of multipolarity $L$ from a
state with angular momentum $I_{i}$ to a state with $I_{f}$ is
given by
\begin{equation}
B(EL;I_{i}\rightarrow I_{f})=\frac{1}{2I_{i}+1}
\sum_{M_{i}M_{f}\mu }\left |\langle \Phi^{\pm}
_{n_{f}I_{f}M_{f}}(\eta,\,\phi)|{\mathcal{M}}_{\mu }(EL)|
\Phi^{\pm}_{n_{i}I_{i}M_{i}}(\eta,\,\phi)\rangle \right |^{2},
\label{transEL}
\end{equation}
where
\begin{eqnarray}
 \Phi^{\pm}_{nIM}(\eta, \, \phi)&=&
 \psi^I_n(\eta ) \varphi^{\pm}(\phi)
 |I0M\rangle \nonumber \\
 &=& \sqrt {\frac {2\Gamma(n+1)}{\Gamma(n+2s+1)}}
 e^{(-a\eta^2/2)} a^s \eta^{2s} L^{2s}_n(a\eta^2)
 \varphi^{\pm}(\phi)
 \sqrt{\frac{2I+1}{32\pi^2}}{D^{I}_{0,\,
 M}}(\theta ).
 \end{eqnarray}

The general form of the multipole operators ${\mathcal{M}}$ in the
collective variables is given in \cite{IG70}.
The electric quadrupole and octupole transition operators for an
axially symmetric nucleus are defined by the deformation variables
$\beta_2$ and $\beta_3$ as
\begin{equation}
{\mathcal{M}}_{\mu }(EL)= M_{L}\beta _{L}{ D_{0\mu }^{L}(\theta
)}, \qquad L=2,3\ ,\qquad (\mu =-L,...,L),
\end{equation}
while the $E1$ (dipole) transition operator is defined as
\cite{DMS86}-\cite{D92}
\begin{equation}
{\mathcal{M}}_{\mu }(E1)=M_{1}\beta _{2}\beta _{3}{ D_{0\mu
}^{1}(\theta )}\ ,\qquad (\mu =0,\pm 1),
\end{equation}
where $M_i$ ($i=1,2,3$) are constants related to the respective
intrinsic moments. In terms of the polar variables $\eta$ and
$\phi$ the above transition operators read
\begin{equation}
{\mathcal{M}}_{\mu }(E1)=M_{1}\frac{\eta ^{2}\cos \phi \sin \phi
}{ \sqrt{d_{2}d_{3}/d^{2}}}{ { D_{0\mu }^{1}(\theta )},}
\end{equation}
\begin{equation}
{\mathcal{M}}_{\mu }(E2)=M_{2}\frac{\eta\cos \phi }{
\sqrt{d_{2}/d}}{ { D_{0\mu }^{2}(\theta )},}
\end{equation}
\begin{equation}
{\mathcal{M}}_{\mu }(E3)=M_{3}\frac{\eta\sin \phi }{
\sqrt{d_{3}/d}}{ { D_{0\mu }^{3}(\theta )}}\ .
\end{equation}

In Eq.~(\ref{transEL}) the integration over the angles $\theta$
involves an integral over three Wigner functions \cite{Ed57},
which leads to the Clebsch-Gordan coefficients $\langle
I_i0L0|I_f0\rangle$. The integration over the variable $\phi$
leads to the following constants
\begin{eqnarray}
I^{++}_{E2}&=&\frac{2}{\pi}\int_{-\pi/2}^{\pi/2}\cos^3\phi d\phi
=\frac{8}{3\pi}, \label{Ippe2}\\
I^{--}_{E2}&=&\frac{2}{\pi}\int_{-\pi/2}^{\pi/2}\cos\phi
\sin^2(2\phi)d\phi =\frac{32}{15\pi}, \label{Imme2}\\
I^{+-}_{E1}&=&\frac{2}{\pi}\int_{-\pi/2}^{\pi/2}\cos^2\phi
\sin\phi\sin(2\phi)d\phi =\frac{16}{15\pi}, \label{Ipme1}\\
I^{+-}_{E3}&=&\frac{2}{\pi}\int_{-\pi/2}^{\pi/2}\cos\phi
\sin\phi\sin(2\phi)d\phi =\frac{1}{2}.
\label{Ipme3}
\end{eqnarray}
The notations $(++)$, $(--)$, and $(+-)$ correspond to the
parities of the functions $\varphi^{\pm}(\phi)$ included in the
integration.

As a result, the reduced $E1$, $E2$, and $E3$
transition probabilities between levels with $|n_iI_i\rangle$ and
$|n_fI_f\rangle$ are given by the expressions
\begin{equation}
B(E1,I_i\rightarrow I_f)=b_{1}\langle I_i010|I_f0\rangle^2
S^2(E1,I_i\rightarrow I_f),
\label{be1}
\end{equation}
\begin{equation}
B(EL,I_i\rightarrow I_f)=b_{L}\langle I_i0L0|I_f0\rangle^2
S^2(EL,I_i\rightarrow I_f)\ ,
\label{beL}
\end{equation}
where
\begin{equation}
S(E1,I_i\rightarrow I_f)=\int_0^{\infty} d\eta
\psi_{n_f}^{I_f}(\eta)\eta^3 \psi_{n_i}^{I_i}(\eta),
\label{se1}
\end{equation}
\begin{equation}
S(EL,I_i\rightarrow I_f)=\int_0^{\infty} d\eta
\psi_{n_f}^{I_f}(\eta)\eta^{2} \psi_{n_i}^{I_i}(\eta),
\label{seL}
\end{equation}
with $L=2,3$. In Eq.~(\ref{beL}) the values of the integrals (\ref{Ippe2}),
(\ref{Imme2}), (\ref{Ipme3}) are included in the constant $b_{L}$.
In Eq.~(\ref{be1}) the constant (\ref{Ipme1}) is included in $b_{1}$.
We remark that if values of different kinds of transition probabilities
are compared, or if branching ratios are considered, the quantities
(\ref{Ippe2})--(\ref{Ipme3}) should be taken into account explicitly.

In the case of transitions between states of the yrast alternating
parity band, $|0\, I_{i}\rangle $ and $|0\, I_{f}\rangle $
(with $n_i=n_f=0$), we obtain the integrals (\ref{se1}) and (\ref{seL})
in the following simple analytic form
\begin{equation}
S(E1,I_{i}\rightarrow I_{f})=\frac{1}{a^2}\frac{\Gamma (s_{i}+s_{f}+2)}
{\sqrt{\Gamma (2s_{i}+ 1)\,\Gamma (2s_{f}+1)}},
\label{se1e}
\end{equation}
\begin{equation}
S(EL,I_{i}\rightarrow I_{f})=\frac{1}{a^{3/2}}\frac{\Gamma
(s_{i}+s_{f}+\frac{3}{2})}{\sqrt{\Gamma (2s_{i}+1)\,\Gamma (2s_{f}+1)}},
\label{seLe}
\end{equation}
where $s_{i}=(1/2)\sqrt{k_{i}^{2}+\widetilde{X}(I_{i})}$,
$s_{f}=(1/2)\sqrt{ k_{f}^{2}+\widetilde{X}(I_{f})}$, and
$a=\sqrt{BC}/\hbar$.

\begin{figure}[t]
\centerline{\epsfxsize=16.cm\epsfbox{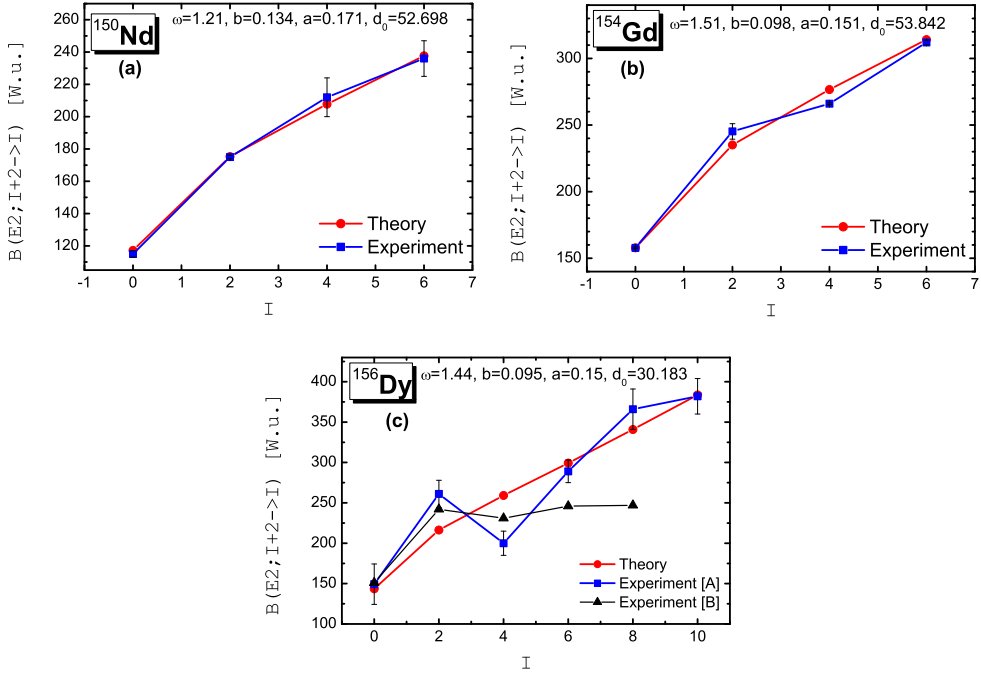}} \caption{(Color
online) Theoretical and experimental B(E2) transition
probabilities (in W.u.) as functions of the angular momentum in
the alternating parity spectra of $^{150}$Nd, $^{154}$Gd and
$^{156}$Dy. The data for $^{150}$Nd, $^{154}$Gd, and for
Experiment[A] in $^{156}$Dy are from \cite{ensdf}; Experiment[B]
in $^{156}$Dy is from \cite{dy156}. The theoretical results are
obtained by Eq. (\protect\ref{beL}). See Sec. 5 for further
discussion.} \label{fig:12}
\end{figure}

In the case of the infinite square well potential (case III), the
model wave function is of the form
\begin{equation}
\Phi^{\pm}_{\nu ,n,
IM}(\eta,\,\phi)=c_{\nu,\,n}J_{\nu}(\kappa_{\nu,\,n}\eta)
\sqrt{\frac{2I+1}{32\pi^2}}{ D^{I}_{0,\,
M}(\theta)}\varphi^{\pm}(\phi)\ ,
\end{equation}
while the integrals over the variable $\eta$ read
\begin{equation}
S(E1,I_i\rightarrow I_f)=\int_0^{\infty} d\eta
J_{\nu_{i}}(\kappa_{\nu_{i},\,n_{i}}\eta)\eta^3
J_{\nu_{f}}(\kappa_{\nu_{f},\,n_{f}}\eta)\ ;
\end{equation}
\begin{equation}
S(EL,I_i\rightarrow I_f)=\int_0^{\infty} d\eta
J_{\nu_{i}}(\kappa_{\nu_{i},\,n_{i}}\eta)\eta^2
J_{\nu_{f}}(\kappa_{\nu_{f},\,n_{f}}\eta).
\end{equation}

\begin{figure}[t]
\centerline{\epsfxsize=16.cm\epsfbox{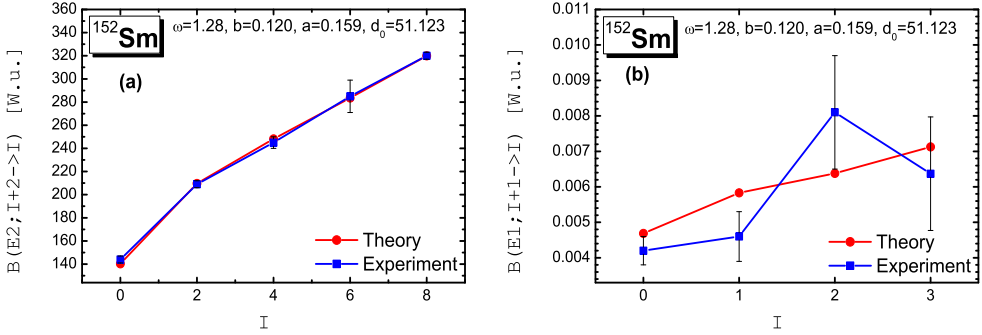}} \caption{(Color
online) Theoretical and experimental B(E2) [part (a)] and B(E1)
[part (b)] transition probabilities (in W.u.) as functions of the
angular momentum in the alternating parity spectrum of $^{152}$Sm
(data from \cite{ensdf}). The theoretical results are obtained by
Eqs. (\protect\ref{be1}) and (\protect\ref{beL}).} \label{fig:13}
\end{figure}

In general, the above formalism can be applied for a detailed analysis
of the electric transition rates in spectra where the collective
quadrupole octupole dynamics carries the characteristics outlined
in the cases I--III of our study. In Figs. 12 and 13 we illustrate
its application to E2 transition probabilities in the nuclei $^{150}$Nd,
$^{152}$Sm, $^{154}$Gd, and $^{156}$Dy, as well as to the E1 transitions in
$^{152}$Sm, in the framework of Case I. The results are obtained with
the parameter sets given in Fig. 7. The quantity $a=\sqrt{BC}/\hbar$,
appearing in Eqs.~(\ref{se1e}) and (\ref{seLe}), has been considered as
a fitting parameter. The constant $b_1$ in Eq.~(\ref{be1}) has been determined
so as to scale the theoretical $E1$ transition values with respect to the
experimental data and takes the value $b_1=1.2\times 10^{-6}$, while the
constants $b_L$ in Eq.~(\ref{beL}) have been set equal to $1$.

We see a good agreement between theory and experiment for the B(E2)
values in $^{150}$Nd [Fig. 12(a)], $^{152}$Sm [Fig. 13(a)], and $^{154}$Gd
[Fig. 12(b)]. In Fig. 12(c), the theoretical E2 transition probabilities in
$^{156}$Dy are compared to two different sets of experimental data,
\cite{ensdf} and \cite{dy156} (with no error bars reported in \cite{dy156}).
We see that the theoretical values follow only the overall increase
of the experimental data. We should, however, remark that the two sets of
data  diverge essentially, especially at the higher angular momenta.
There is also some discrepancy between theory and experiment in
the E1 transition values in $^{152}$Sm [Fig. 13(b)]. The results in
Figs. 12(c) and 13(b) suggest that further examination of the
formalism, as well as of the experimental data, may be necessary.
The further analysis of data on electric transitions in a wider
range of nuclei will be the subject of future work.

\section{Influence of the $\gamma$ deformation mode on the
$\beta_2$-$\beta_3$ collective motion}

As it has been mentioned in Sec. 4, the present model framework
does not include the $\gamma$ degree of freedom. Here we briefly
discuss the possible ways in which this can be done and shortly
estimate the influence of the $\gamma$ deformation mode on the
collective motion in the $\beta_2$-$\beta_3$ space. The rotation
energy of a system with presence of axial and triaxial quadrupole
modes ($\beta_2$ and $\gamma$ ) and axial octupole degree of
freedom ($\beta_3$) can be given by \cite{JPDav68,Maslov06}
\begin{equation}
\hat{T}_{ \mbox{\scriptsize rot}}
=\frac{1}{2}\sum_{i=1}^{3}\frac{\hat{I}_i^2}{J_{i}^{(2)}+J_{i}^{(3)}},
\label{roten}
\end{equation}
where
\begin{equation}
J_{i}^{(2)}=4B_{2}\beta_2^{2}\sin^{2}\left(\gamma-\frac{2\pi
i}{3}\right), \qquad (i=1,2,3)\label{qmomin}
\end{equation}
are the moment-of-inertia components of the quadrupole shape about
the axes 1, 2 and 3, while those of the axially symmetric octupole
shape are
\begin{equation}
J_{1}^{(3)}=J_{2}^{(3)}=6B_{3}\beta_{3}^{2},\qquad J_{3}^{(3)}=0.
\end{equation}

A simple estimation of the $\gamma$- influence can be done by
assuming small variations of the system around $\gamma =0$, as in
the case of the X(5) model \cite{IacX5}. Then the quadrupole
moment-of-inertia components (\ref{qmomin}) can be taken as
\begin{equation}
J_{1}^{(2)}=J_{2}^{(2)}=3B_{2}\beta_{2}^{2},\qquad
J_{3}^{(2)}=4B_{2}\beta_{2}^{2}\sin^{2}\gamma.
\end{equation}
As a result the rotation energy (\ref{roten}) obtains the form
\begin{eqnarray}
\hat{T}_{\mbox{\scriptsize rot}} =\frac{1}{2}\left (
\frac{\hat{I}^{2}-\hat{I}_{3}^{2}}{3B_{2}
\beta_{2}^{2}+6B_{3}\beta_{3}^{2}}+
\frac{\hat{I}_{3}^{2}}{4B_{2}\beta_{2}^{2}\sin^{2}\gamma}\right ).
\label{roten1}
\end{eqnarray}
The first term in (\ref{roten1}) corresponds to the centrifugal
term in the quadrupole-octupole potential (\ref{Ub2b3I}). The
second term in (\ref{roten1}) provides the influence of the
$\gamma$ mode on the potential. After taking into account
Eq.~(\ref{roten1}) with $d_2=3B_2$ and $d_3=6B_3$, and by
including a $\gamma$- oscillation term, the potential
(\ref{Ub2b3I}) can be generalized in the form
\begin{equation}\label{Ub2b3gam}
U(\beta_2,\beta_3,\gamma ,I)=\frac{1}{2}C_2{\beta_2}^{2}+
\frac{1}{2}C_3{\beta_3}^{2} +\frac{1}{2}C'_2{\gamma}^{2}
+\frac{X(I)-K^2/2}{d_2\beta_2^2+d_3\beta_3^2}+
\frac{3K^{2}/2}{4d_{2}\beta_{2}^{2}\sin^{2}\gamma},
\end{equation}
where $K$ is the projection of the angular momentum on the
body-fixed $z$-axis. Then for a fixed value of $\gamma$ the
extremum conditions (3) and (4) in Sec. 2 provide the following
cases for the bottom of the potential (\ref{Ub2b3gam}) in the
$\beta_2$-$\beta_3$ space.

1) $\beta_{3\mbox{\scriptsize min}}=0$ with
\begin{eqnarray}
\beta_{2\mbox{\scriptsize min}}=\pm\left\{ \frac{2}{d_2C_2} \left
[ X(I)+\frac{K^2}{2}\left (\frac{3}{4\sin^{2}\gamma }-1\right )
\right ] \right\}^{1/4}. \label{b2min}
\end{eqnarray}
\smallskip

2) $\beta_{2\mbox{\scriptsize min}}\neq0$ and
$\beta_{3\mbox{\scriptsize min}}\neq0$ with the condition
\begin{eqnarray}
C_2=\frac{[2X(I)-K^2]d_2}{(d_2\beta_{2\mbox{\scriptsize
min}}^2+d_3\beta_{3\mbox{\scriptsize min}}^2 )^2}
+\frac{3K^2}{4d_{2}\beta_{2\mbox{\scriptsize
min}}^{4}\sin^{2}\gamma} \ \ \mbox{and}\ \ C_3=\frac{[2X(I)-K^2
]d_3}{(d_2\beta_{2\mbox{\scriptsize
min}}^2+d_3\beta_{3\mbox{\scriptsize min}}^2)^2}\ . \label{CLgam}
\end{eqnarray}

The following comments can be done on the above result.

i) The appearance of $\beta_{2}^{2}$ in the denominator of the
second term in (\ref{Ub2b3gam}) divides the $\beta_2$-$\beta_3$
space into two half-spaces, $\beta_2>0$ and $\beta_2<0$, separated
by an infinite potential barrier at $\beta_2=0$. For this reason
the potential minimum with $\beta_{2\mbox{\scriptsize min}}=0$ and
$\beta_{3\mbox{\scriptsize min}}\neq 0$ does not appear.

ii) Eqs.~(\ref{b2min}) and (\ref{CLgam}) illustrate the ways in
which the term involving the $\gamma$ deformation mode can shift
the position of the potential minima in the $\beta_2$-$\beta_3$
space (compare with cases i)--iii) including Eq.~(\ref{CL}) in
Sec. 2). Note that for $K=0$ the influence of the $\gamma$ mode on
the $\beta_2$-$\beta_3$ potential shape automatically disappears.
This is a limit in which the $\beta_2$ and $\gamma$ degrees of
freedom are weakly coupled and can be adiabatically separated,
{\em which is implied in the framework of the present work}. The
involvement of the $K=2$ configurations in the collective motion
implies the consideration of a strong $\beta_2$-$\gamma$ coupling
\cite{Caprio05}.

iii) The involvement of the $\gamma$ degree of freedom in the
above way would influence the correlation between the axial
$\beta_2$ and $\beta_3$ variables due to the appearance of the
second term in $C_2$ [Eq.~(\ref{CLgam})], as a consequence of the
last term in the potential (\ref{Ub2b3gam}). Now, in terms of the
polar variables, the potential will depend on both $\eta$ and
$\phi$, so that the variables in the Schr\"{o}dinger equation
cannot be directly separated. This could be done in a way similar
to the adiabatic separation of the $\beta$ and $\gamma$ degrees of
freedom in the X(5) model framework \cite{IacX5}, as well as in
the framework of the AQOA model \cite{AQOA}. Alternatively, the
problem could be solved numerically in a way similar to the
approach of Ref.~\cite{Caprio05}.

A more general way to examine the influence of the $\gamma$
deformation mode on the  qua\-dru\-pole-octupole motion of the
system could be based on the complete form (\ref{qmomin}) of the
quadrupole moment-of-inertia components, so that the $\gamma$
variable would not be limited in the vicinity of zero.
Furthermore, non-axiality of the octupole degree of freedom can be
considered. Any efforts in these directions should be based on
numerical solution of the problem.

\section{Summary and conclusions}

The present study outlines some dynamical properties
of a system with {\em simultaneously} manifesting quadrupole and
octupole degrees of freedom. We remark that the obtained results
represent a restricted class of {\em exact analytic} solutions
of the problem. This is due to the correlation (\ref{dbratio})
between the mass and the inertial parameters, which essentially
simplifies the Hamiltonian (\ref{Hqo01}) in the form of
Eq.~(\ref{Hqo02}). In addition, the correlation (\ref{dcratio}) between
the inertial and oscillator parameters brings the potential in a
form depending on the ``effective deformation'' variable $\eta$
only, and not on the relative ``angular'' variable $\phi$, thus
allowing an {\em exact} separation of variables in the Schr\"{o}dinger
equation. As it is explained in Sec. 4, the above correlations
provide a {\em coherent interplay} between the quadrupole and
octupole collective modes. In this respect, the presently considered
potentials suggest some specific properties of quadrupole--octupole
collectivity, which can be developed in various nuclear regions.

However, despite the above limitations (the necessary price we
pay for solving the problem exactly) we were able to identify a
region of nuclei where the assumed ``equal'' presence of quadrupole
and octupole degrees of freedom can take place in the collective
motion. We found that the structure of the  spectrum in the
case of the potential (\ref{cpotenc}), illustrated in Fig.~6, is
similar to the structure of alternating parity bands in some rare
earth nuclei. On this basis, we have reproduced quite accurately the
energy levels and the staggering patterns in the nuclei $^{150}$Nd,
$^{152}$Sm, $^{154}$Gd, and $^{156}$Dy, as demonstrated in Figs.~7
and 8. In these spectra the reduction of the  staggering
amplitude indicates the {\em trend of forming octupole
deformations} towards the higher angular momenta. However, the slow
decrease of the parity effect does not allow this to happen at
reasonable (observed) angular momenta. The B(E2) transition
probabilities have been also described with a reasonable accuracy
(Figs. 12 and 13), while the result for B(E1) transition probabilities
in $^{152}$Sm [Fig. 13(b)] suggests further tests of the formalism and
analysis of additional experimental data.

We remark that the energy expression (\ref{spect1a}) cannot
reproduce the complicated beat staggering effects observed in the
octupole bands of light actinide nuclei \cite{DBoct00}. The latter
have been described \cite{MYDS06}, with good accuracy, by the use
of the Quadrupole--Octupole Rotation Model \cite{octahed01}. Thus,
compared to the light actinide region, the application of
expression (\ref{spect1a}) indicates a different behavior of the
quadrupole--octupole collectivity in the rare earth nuclei, with a
{\em less developed octupole deformation} and a {\em more strongly
pronounced octupole vibration mode}. The present analysis suggests
that in this case the {\em coherent (equal) contribution} of
quadrupole and octupole oscillations can take place in the
collective motion of nuclei.

The potentials with fixed energy minima (cases II and III) can be
related generally to a situation in which the vibrational and
rotational degrees of freedom are weakly coupled. Then the rotational
angular momentum slightly affects the quadrupole-octupole
vibration motion, which suggests a constant (or nearly constant)
behavior of the staggering amplitude. In particular this is well
seen in the case II.B, where the direct contribution of the rotational
motion is excluded. Thus, the spectrum illustrated in Fig.~10 suggests
an essentially quadrupole-octupole {\em vibrational} motion of the
system. Cases II.A and III, in which the rotational mode is taken
into account, suggest quadrupole-octupole vibrations with an
adiabatically manifested {\em rotational} motion. We remark that
the constant staggering patterns, illustrated in Figs. 9 and 10,
are in some meaning idealistic cases, as far as the current
experimental data do not show such a strong persistence of the
parity effect at high angular momenta. On the other hand, the square
well potential of case III appears to be applicable to
examining the possible critical behavior of the quadrupole--octupole
collectivity in different nuclear regions. Studies in this direction
have been implemented recently in the light actinide nuclei
\cite{AQOA}. We suggest that further analysis of experimental data
for quadrupole--octupole spectra would be of use for testing the
prediction of the staggering pattern illustrated in Fig. 11 for
the case of the square well potential (\ref{sqwell}).

It is important to note that the present exactly solvable model
can be naturally extended, beyond the ``coherent interplay''
assumption, to a more general non-analytic problem in the
following two ways. First, we can  release the correlation
(\ref{dcratio}) between the inertial and oscillator parameters,
allowing the potential to depend on the variable $\phi$. Then the
problem can still be transformed into a form having an analytical solution,
by performing an ``approximate'' separation of variables, as done in
\cite{AQOA} and in the framework of the X(5) symmetry model
\cite{IacX5}. The second extension would be to release the
correlation (\ref{dbratio}) between the mass and inertial parameters.
This would allow us to examine different ways in which the coupled
quadrupole and octupole degrees of freedom enter the collective motion.
In this case, however, more sophisticated mathematical and numerical
techniques have to be sought to solve the problem.

Finally, we also remark that the developed formalism contains
several limits. Thus, when the quadrupole variable is frozen to
some stable quadrupole deformation, the potentials and the spectra
of the cases I-III transform to the respective ones appearing in
the one-dimensional problem \cite{MYDS06,Bizz05}. Another
interesting limit can be obtained by appropriate parameter values,
for which the difference $\Delta k^2$ is negligible compared to
$\widetilde{X}(I)$ for all angular momentum values. Then the
staggering effect vanishes, and the odd and even angular momentum
sequences appear in a single non-perturbed collective band. For
example, if such a transition is performed in case III, the
spectrum presented in Fig.~11 is reduced to the structure supposed
to correspond to the transition between octupole vibrations and
stable octupole deformation, in which a single octupole band is
formed \cite{AQOA}. It is also of interest to take into account
the non-axiality of the quadrupole and/or the octupole degree of
freedom, in order to examine how the present results are modified.
Studies in these directions will be the subject of further work.
\bigskip

\noindent {\Large \bf Acknowledgments}
\medskip

\noindent We thank Prof. P. G. Bizzeti and Prof. R. V. Jolos for
valuable discussions and comments. This work is supported by DFG and
by the Bulgarian Scientific Fund under contract F-1502/05.

%\newpage

\end{document}